\documentstyle[pra,epsf,aps]{revtex}
\begin{document}
\title{{Critical Exponents from
Seven-Loop Strong-Coupling $\phi^4$-Theory in Three Dimensions
}}
\author{Hagen Kleinert%
 \thanks{Email: kleinert@physik.fu-berlin.de ~~ URL:
http://www.physik.fu-berlin.de/\~{}kleinert ~~ Phone/Fax:
 0049 30 8383034 }}
\address{Institut f\"ur Theoretische Physik,\\
Freie Universit\"at Berlin, Arnimallee 14,
14195 Berlin, Germany}
\maketitle
\begin{abstract}
Using strong-coupling quantum field theory
we calculate highly accurate critical exponents $ \nu ,  \eta $
from  new seven-loop expansions
in three dimensions. Our theoretical
value for the critical exponent $ \alpha $ of the specific heat
near the $ \lambda $-point
of superfluid helium is $ \alpha =-0.01294\pm0.00060$,
in excellent agreement
with the space shuttle experimental value $ \alpha =-0.01285\pm0.00038$.
\end{abstract}

\pacs{}
%
\noindent
{\bf 1.} The accurate calculation
of critical  exponents from field theory
presents a theoretical challenge since the relevant
information is available only from
divergent power series expansions.
The results
are also of practical relevance
since they
predict the
the outcome many possible future experiments
on many second-order phase transitions.
In recent work \cite{kl}
we have developed a novel method
for extracting these exponents
from such expansions via a
 strong-coupling theory
of scalar fields
with a $\phi^4$-interaction.
The fields are assumed to have
$n$ components with an action which is
O($n$)-symmetric.
As an application, we have
used  available six-loop perturbation expansions
of the renormalization constants in three dimensions
 \cite{1,3,anton}
 to calculate
the critical exponents
for
all
 O($n$) universality classes with high precision.
Strong-coupling theory
works also in $4- \epsilon $ dimensions \cite{klep},
and is capable of interpolating
between the expansions
in $4- \epsilon $ with those in $2+\epsilon$
dimensions of the nonlinear $ \sigma $-model
\cite{klepp}.

The purpose of this note is to
improve significantly the accuracy
of our earlier results in three dimensions \cite{kl}
by
making use of new seven-loop expansion
coefficients for
the critical exponents $ \nu $ and $  \eta $ \cite{MN}
and, most importantly, by
applying a more powerful extrapolation method
to infinite order than before.
The latter makes
our results as
accurate as those obtained
by Guida and Zinn-Justin \cite{GZ}
via a more sophisticated resummation technique
based on analytic mapping and Borel transformations,
which in addition takes into account
information on the
large-order growth of the expansion coefficients.
We reach this accuracy without
using that information which, as we shall
demonstrate
at the end in Section~5,
has practically no influence on the results,
except
for lowering $ \omega $ slightly (by less than $\sim 0.2 \%$).
The reason for the
little importance
of the large-order information
in our approach
is that
 the critical exponents are obtained from evaluations of expansions
 at infinite bare couplings.
The information on the large-order behavior, on the other hand,
specifies
the  discontinuity at the
tip of the left-hand cut which starts at the origin
of the complex-coupling constant plane \cite{PI}.
This is
too far from the infinite-coupling limit to be of relevance.
In our resummation scheme
for expansions  in powers of the bare coupling constant ,
an important role is played by the critical exponent of approach
to scaling $ \omega $, whose precise calculation
by the same scheme
is crucial for obtaining high accuracies in all other critical exponents                                                                                                                                                                                                                                                              .
It is determined by the condition that the renormalized coupling strength
$g$
goes against a constant $g^*$ in the strong-coupling limit.
The knowledge of
 $ \omega $
is
more yielding
than the large-order information
in previous resummation schemes in which
the critical exponents are determined
as a function of the renormalized coupling constant $g$ near $g^*$
which is of order unity,
thus lying a finite distance
away from the left-hand cut in the complex $g$-plane.
Although
these determinations are sensitive
to the discontinuity at the tip of the cut,
it must be realized that
the influence of the cut is very small
due to the smallness of the fugacity of the leading instanton,
which carries a Boltzmann factor $e^{-{\rm const}/g}$.
\\

\noindent
{\bf 2.}
We briefly recall the
available
expansions \cite{anton} of the renormalized coupling ${\bar g}\equiv g/m$
in terms of the bare coupling
$\bar g_0\equiv g_0/m$ for all O($n$):
\begin{eqnarray}
{\bar g}/{\bar g}_0&=&1 \!-\! {\bar g_0}\,\left( 8\!+\!n \right)  \!+\!
  {{\bar g_0}^{2}}\,\left(2108/27 \!+\! 514n/27 \!+\! {n^{2}} \right) \nonumber \\&&  \!+\!
  {{\bar g_0}^{3}}\,\left( \!-\!878.7937193 \!-\! 312.63444671n \!-\! 32.54841303{n^{2}}
 \!-\!
     { n^{3}} \right) \nonumber \\&&  \!+\! {\bar g_0^{4}}\,
   \left( 11068.06183 \!+\! 5100.403285n \!+\! 786.3665699{n^{2}} \!+\!
     48.21386744{n^{3}} \!+\! {n^{4}} \right)  \nonumber \\&& \!+\!
  {{\bar g_0}^{5}}\,\left( \!-\!153102.85023 \!-\! 85611.91996n \!-\! 17317.702545{n^{2}} \!-\!
     1585.1141894{n^{3}} \!-\! 65.82036203{n^{4}} \!-\! {n^{5}} \right)
  \nonumber \\&& \!+\!
 {{{\bar g_0}}^6}\,\left( 2297647.148\, \!+\! 1495703.313\,n \!+\!
     371103.0896{n^2} \!+\! 44914.04818{n^3}
\right. \nonumber \\&& \left.~~~~~~~\!+\! 2797.291579{n^4} \!+\!
     85.21310501{n^5} \!+\! {n^6} \right),
\label{gfg-0}\end{eqnarray}
and of the critical exponents \cite{not}
\begin{eqnarray}
\!\!\omega ({\bar g}_0) & = &
\!-\!1 \!+\!2 {\bar g_0}\,\left( 8 \!+\! n \right) \!-\!
  {{{\bar g_0}}^{2}}\,\left( 1912/9 \!+\!452n/9\!+\!2{n^{2}} \right) \nonumber \\&&   \!+\! {{{\bar g_0}}^{3}}\,\left( 3398.857964 \!+\! 1140.946693n \!+\!
     95.9142896{n^{2}} \!+\! 2{n^{3}} \right)  \nonumber \\&&  \!+\!
  {{{\bar g_0}}^{4}}\,\left( \!-\!60977.50127 \!-\! 26020.14956n \!-\!
     3352.610678{n^{2}} \!-\! 151.1725764{n^{3}} \!-\! 2{n^{4}} \right)
 \nonumber \\&&   \!+\!
  {{{\bar g_0}}^{5}}\,\left( 1189133.101 \!+\! 607809.998n \!+\!
     104619.0281{n^{2}} \!+\! 7450.143951{n^{3}} \!+\! 214.8857494{n^{4}} \!+\!
     2{n^{5}} \right)  \nonumber \\&&  \!+\! {{{\bar g_0}}^{6}}\,
   \left( \!-\!24790569.76 \!-\! 14625241.87n \!-\!
     3119527.967{n^{2}}\right.
 \nonumber \\&&\left.~~~~~~~~~\,  \!-\! 304229.0255{n^{3}} \!-\!
     14062.53135{n^{4}} \!-\! 286.3003674{n^{5}} \!-\! 2{n^{6}} \right),
\\
\!\!\eta({\bar g})&=&
{{{\bar g_0}}^2}\,\left( 16/27 \!+\! 8n/27 \right)  \!+\!
  {{{\bar g_0}}^3}\,\left( \!-\!9.086537459 \!-\! 5.679085912n \!-\!
     0.5679085912{n^2} \right)   \nonumber \\&&  \!+\!
  {{{\bar g_0}}^4}\,\left( 127.4916153 \!+\! 94.77320534n \!+\! 17.1347755{n^2} \!+\!
     0.8105383221{n^3} \right)   \nonumber \\&&  \!+\!
  {{{\bar g_0}}^5}\,\left( \!-\!1843.49199 \!-\! 1576.46676n \!-\! 395.2678358{n^2} \!-\!
     36.00660242{n^3} \!-\! 1.026437849{n^4} \right)               ,
 \label{indg-0} \nonumber \\&&    \!+\!
  {{{\bar g_0}}^6}\,\left( 28108.60398 \!+\! 26995.87962n \!+\! 8461.481806{n^2} \!+\!
     1116.246863{n^3} \!+\! 62.8879068{n^4} \!+\! 1.218861532{n^5} \right),
\label{@eta}\label{indg-0}\\
\!\!\eta_m({\bar g})&=&
{\bar g_0}\,\left( 2 \!+\! n \right)  \!+\!
  {{{\bar g_0}}^2}\,\left( \!-\!523/27 \!-\! 316n/27 \!-\! {n^2} \right)
 \nonumber \\&&
  \!+\!
  {{{\bar g_0}}^3}\,\left( 229.3744544 \!+\! 162.8474234n \!+\! 26.08009809{n^2}
\!+\!
     {n^3} \right)
 \nonumber \\&&
 \!+\! {{{\bar g_0}}^4}\,
   \left( \!-\!3090.996037 \!-\! 2520.848751n \!-\! 572.3282893{n^2} \!-\!
     44.32646141{n^3} \!-\!{n^4} \right)
  \nonumber \\&&  \!+\!
  {{{\bar g_0}}^5}\,\left( 45970.71839 \!+\! 42170.32707n \!+\! 12152.70675{n^2} \!+\!
     1408.064008{n^3} \!+\! 65.97630108{n^4} \!+\! {n^5} \right)
 \nonumber \\&& \!+\!
  {{{\bar g_0}}^6}\,\left( \!-\!740843.1985 \!-\! 751333.064n \!-\! 258945.0037{n^2}
\!-\!
    39575.57037{n^3} \!-\! 2842.8966{n^4} \!-\! 90.7145582{n^5} \!-\! {n^6} \right)
,
\label{@gammam}\end{eqnarray}
where $\eta_m\equiv2- \nu ^{-1} $.
To save space we have omitted a factor $1/(n+8)^n $
accompanying each power $\bar g_0^n$ on the right-hand sides.
The additional seventh-order coefficients
have been calculated for
$n=0,\,1,\,2,\,3$
and are \cite{MN} (these without a factor $1/(n+8)^7$ on the right-hand side)
\begin{eqnarray}
 \eta^{(7)}
 =\left\{
\begin{array}{r}
- 0.2164239372\\
- 0.2395467913\\
- 0.2414247646\\
- 0.2333645418
\end{array}\right\}
\bar g_0^7,~~~~~
 \nu^{-1\,(7)} =\left\{
\begin{array}{r}
-6.0998295658\\
-7.0482198342\\
-7.3780809849\\
-7.3808485089
\end{array}\right\}
\bar g_0^{7}~~~~~
{{\rm for}}       ~~~~~
\left\{ \begin{array}{l}
n=0\\
n=1\\
n=2\\
n=3
\end{array}~%
\right\} .
\label{@sev}\end{eqnarray}

\begin{figure}[tbhp]
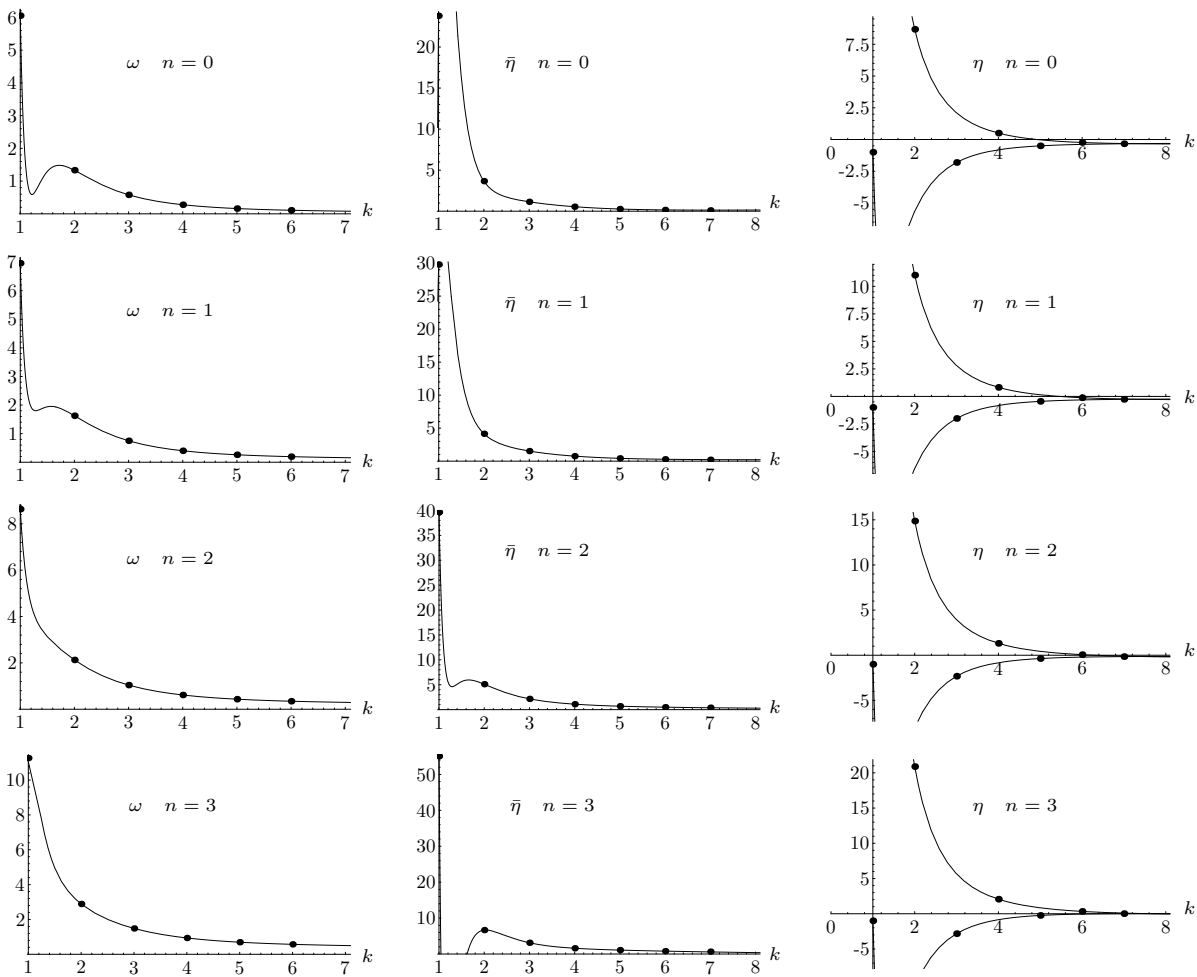

~\\
~~~~~~\input omeasr0.tps \\
~~~~~~\input omeasr1.tps  \\
~~~~~~\input omeasr2.tps    \\
~~~~~~\input omeasr3.tps      \\
\caption[Precocity of
large-order behavior of coefficients of
 the expansions of the critical exponents
$ \omega $, $  \bar \eta\equiv \nu^{-1}+ \eta -2 $, and $ \eta $
in powers of the
renormalized
coupling constant]{Precocity of
large-order behavior of coefficients of
 the expansions of the critical exponents
$ \omega $, $  \bar \eta\equiv \nu^{-1}+ \eta -2 $, and $ \eta $
in powers of the
renormalized
coupling constant.
The dots show the relative deviations exact/asymptotic-1.
The curves are plots of
the asymptotic expressions in Eqs.~(\ref{@omas})--(\ref{@etpas})
listed in Table \ref{Table asym}.
The curve for $ \omega $ is the smoothest, promising
the best extrapolation to the next orders, with  consequences
to be discussed in
Section 5.
}
\label{lo0}\end{figure}

\noindent
{\bf 3.} It is instructive to see how close the new coefficients
are to their large-order
limiting values derived from
instanton calculations, according to which
the expansion coefficients with respect
to the renormalized
coupling $\bar g$
should grow for large order $k$
as follows
\cite{parisi}:
\begin{eqnarray} \label{@omas}
    \omega  ^{(k)} & = & \gamma_  \omega   (-a)^k k! k   \Gamma(k+b_  \omega
)
\left(1+\frac{c_ \beta ^{(1)}}{k}        +\frac{c_ \beta ^{(2)}}{k^2}+\dots\right),
 \\
   \eta ^{(k)} & = & \gamma _\eta (-a)^k k! k \Gamma(k+b_\eta)
\left(1+\frac{c_ \eta  ^{(1)}}{k}        +\frac{c_  \eta  ^{(2)}}{k^2}+\dots\right),
\label{@etas}
\\
    \bar{\eta}^{(k)} & = & \gamma _{\bar{\eta}} (-a)^k
		   k! k  \Gamma (k+b_{ \bar{\eta}})
\left(1+\frac{c_ {\bar{\eta}} ^{(1)}}{k}        +\frac{c_ { \bar{\eta}}  ^{(2)}}{k^2}+\dots\right).
\label{@etpas}
\label{@grow}\end{eqnarray}
where $ \bar{\eta}\equiv \eta + \nu ^{-1}-2$.
The growth parameter $a$ proportional to the inverse euclidean action
of the classical {\em instanton\/} solution $\varphi_c  ({\bf x})$
to the field equations
\begin{eqnarray} 
   a =
(D-1)\frac{16\pi}{I_4}\frac{1}{N+8}=0.14777423 \frac{9}{N+8}.
\end{eqnarray}
The quantity $I_4$ denotes the integral
 $I_4= \int d^D  x [\varphi_c  ({\bf x})]^4$.
Its numerical values in two and three dimensions $D$
are listed in Table \ref{Table 6}. The growth parameters $b_\beta,b_ \eta , b_{ \bar{\eta}}  $,
are directly related to
the number of zero-modes in the fluctuation determinant around
the instanton, which is $D+N$ (associated with $D$ translations,
$n-1$ rotations, and one dilation).
Their values are
\begin{eqnarray}
  b_ \omega =b_\beta+1  = \frac{1}{2} (D+5+n) , ~~~
      b _\eta = \frac{1}{2} (D+1+n),~~~
      b _{\bar{\eta}} = \frac{1}{2} (D+3+n)
\end{eqnarray}
The prefactors $\gamma_ \beta ,\gamma_ \eta , \gamma_{ \bar{\eta}}  $
in
(\ref{@omas})--(\ref{@grow})
require the calculation of the full
fluctuation determinants. This yields
\begin{eqnarray} 
  \gamma_ \beta   & \equiv & \frac{(n+8)2^{(n+D-5)/2} 3^{-3(D-2)/2}}
		  {\pi ^{3+D/2}\Gamma \left( 2+ \frac{1}{2}n\right) }
		  \left( \frac{I_1^2}{I_4}\right) ^2
\left( \frac{I_6}{I_4}-1\right) ^{D/2}
                 D_L^{-1/2} D_T^{-(n-1)/2} e^{-1/a}.
\end{eqnarray}
The constants $I_1$, $I_2$, $I_6$ are
 generalizations
of the above integral $I_4$:
$I_p= \int d^D  x [\varphi_c  ({\bf x})]^p$,
and $D_L$  and  $D_T$ are found
from the longitudinal
and transverse parts of the fluctuation determinants.
Their numerical values are given in Table \ref{Table 6}.
The constant $ \gamma _ \beta $ is the prefactor of growth in the
expansion coefficients
of the $ \beta $-function (the integral over $ \omega (\bar g))$:
 $ \beta ^{(k)}\approx
 \gamma_  \beta    (-a)^k k!    \Gamma(k+b_ \beta)
$.
The prefactors in
$\gamma_  \omega $,
$\gamma_ \eta$,
and $ \gamma_{ \bar{\eta}}  $ in (\ref{@omas})--(\ref{@etpas})
are related to $\gamma_  \beta$ by
\begin{eqnarray}
      \gamma _ \omega   =-a  \gamma _{ \beta },~~~~ ~
      \gamma _\eta  =  \gamma _{\bar{\eta}}
		 \frac{2H_3}{I_1 D(4-D)}    ,~~~~~
   \gamma _{\bar{\eta}}  =  \gamma _ \beta \frac{n+2}{n+8} (D-1)
		  4\pi \frac{I_2}{I_1^2}.
\end{eqnarray}
where $I_2 = (1-D/4)I_4$ and $H_3$ are listed in
Table \ref{Table 6}.
The numerical values of all growth
parameters for are  listed in Table~\ref{compiled}.
In Figs.~\ref{lo0} we show a comparison between the exact coefficients
and their asymptotic forms (\ref{@grow}).

~\\
\noindent
{\bf 4.} The critical exponents are  derived
from
the divergent expansions
(\ref{gfg-0})--(\ref{@sev})
by going to the
limit $\bar g_0\rightarrow \infty$.
In a theory with scaling behavior, the renormalized coupling constant
$\bar g$ tends to a limiting value $\bar g^*$
as follows:
\begin{equation}
\bar g(\bar g_0)=\bar g^*-\frac{{\rm const}}{\bar g_0^{  \omega/ \epsilon}}+\dots~,
\label{appr}\end{equation}
where $g^*$ is commonly referred to as the infrared-stable
fixed point, and $ \omega$ is called the critical
exponent of the approach to scaling.
The same exponent governs the approach
to scaling of every function $G(\bar g)$
which behaves like
%
$G(\bar g)=G(\bar g^*)+G'(\bar g^*)\times   {{\rm const}}/{\bar g_0^{  \omega}}+\dots~.$
%

How do we recover the $\bar g_0\rightarrow \infty$ -limits
of a function $f(\bar g_0)$
if we know the first $N$
terms  of     its asymptotic
expansion
$ f_N(\bar g_0)=\sum_{n=0}^N  a_n  \bar g_0  ^n$?
Extending systematically the behavior (\ref{appr})
we
shall assume that
$ f(\bar g_0)$  approaches its constant limiting value $f^*$
in the form of an inverse power series
\cite{confl}
 $f_M(\bar g_0)=  \sum_{m=0}^M b_m  (\bar g_0 ^{- \omega }) ^m$.
This strong-coupling expansion
has usually a finite convergence radius $g_s$
(see \cite{kl,PI,int}).
The $N$th approximation to the
value $f^*$ is obtained from the formula
\begin{eqnarray}
&&f_N^* =\mathop{\rm opt}_{\hat{g}_0}\left[
\sum_{j=0}^N a_{j}^{\rm } \hat{g}_0^j
 \sum_{k=0}^{N-j}
      \left( \begin{array}{c}
	      -  j/ \omega  \\ k
	     \end{array}
      \right)
     (-1)^{k}   \right] ,
\label{coeffb}\end{eqnarray}
where the expression in brackets
has to be optimized in the variational parameter
$\hat g_0$.
The optimum is
the smoothest among all real extrema. If there are no such extrema,
which happens for the even approximants,
the
turning points serve the same purpose.

From the theory \cite{kl}, we expect the
exact values
to be approached exponentially fast  with the order $N$ of the available
expansions, with the error decreasing like
$e^{-cN^{1- \omega }}$.
In order to extrapolate
our results to $N=\infty$, we
plot
the data
against the variables $x_N=e^{-cN^{1- \omega }}$ (see also Addendum to
Ref.~\cite{kl}).
This is done separately for even and odd approximants, since
the former stem from extrema, the latter from turning points.
The unknown constants $c$
$c$ are determined by fitting to each set of points a slightly curved parabola
and making them intersect the vertical axis
at the same point, which
yields the extrapolated critical exponent
listed on top of each figure (together with the seventh-order value in parentheses, and the optimal
parameter $c$).

\begin{figure}[tbhp]
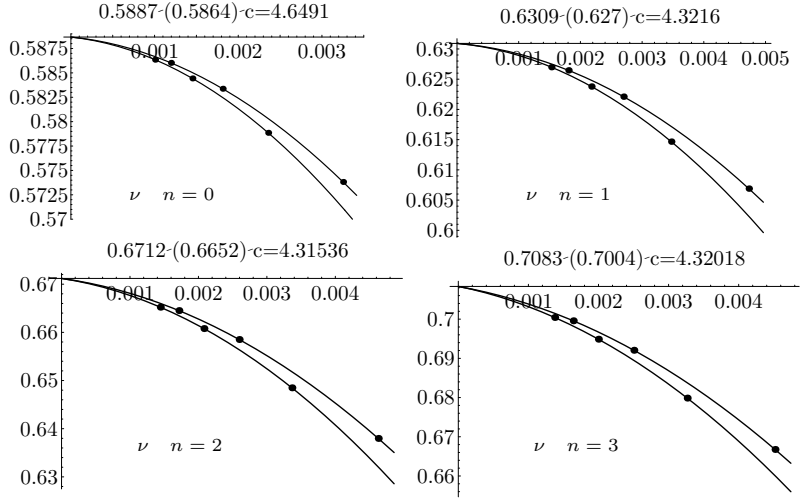

~\\
\input crnu0g.tps  ~\\~\\
\caption[]{Strong-coupling values
for the critical exponent $ \nu ^{-1}$
obtained from expansion (\protect\ref{@gammam})
via formula (\ref{coeffb}),
for increasing orders
$N=2,3,\dots,7$
 of the approximation.
The exponents are plotted against the variable $x_N=e^{-cN^{1- \omega }}$
and should
for large $n$ lie on a straight line.
Here at finite $N$, even and odd approximants
may be connected by slightly curved parabolas
whose common intersection with the vartical exis determines the critical exponents
for $N=\infty$. More details on the determination of the
constant $c$ are given in the text.
The numbers on top give the extrapolated critical exponents
and, in parentheses, the highest approximants, to illustrate
the extrapolation distance.
}
\label{nu0}\end{figure}
Following this procedure, we find from the expansion
(\ref{@gammam}) for $ \nu ^{-1}$
the approximants $ \nu_N ^{-1}$
via formula (\ref{coeffb}).
Extrapolating
separately even and odd approximants $ \nu _N$,
we
determine
the limiting value $ \nu $, as shown in Fig.~\ref{nu0}.
The $ \omega $-values used for this extrapolation
are those of
Ref.~\cite{klep}, listed in the last colum of Table \ref{TableI}:
\begin{equation}
\omega_6
 =\left\{
\begin{array}{r}
0.810\\ 0.805\\ 0.797\\ 0.790
\end{array}\right\}
~~~~~
{{\rm for}}       ~~~~~
\left\{ \begin{array}{l}
n=0\\
n=1\\
n=2\\
n=3
\end{array}\right\}.~~ %
\label{@omex}\end{equation}
They lead to the $ \nu $-values
$\nu_7=\left\{
0.5883,
0.6305,
0.6710,
0.7075\right\} $, the entries in this vector referring to $n=0,1,2,3$.
Since these results depend
on the critical exponents $ \omega $,
it is useful to study the dependence of the extrapolation on $ \omega $,
with the result
\begin{equation}
  \nu_7
 =\left\{
\begin{array}{r}
0.5883+0.0417\times ( \omega -0.810)\\
0.6305+0.0400\times ( \omega -0.805)\\
0.6710+0.0553\times ( \omega -0.800)\\
0.7075+0.1891\times ( \omega -0.797)
\end{array}\right\}
~~~~~
{{\rm for}}       ~~~~~
\left\{ \begin{array}{l}
N=0\\
N=1\\
N=2\\
N=3
\end{array}\right\} .%
\label{@nuexdel}\end{equation}

\begin{figure}[tbhp]
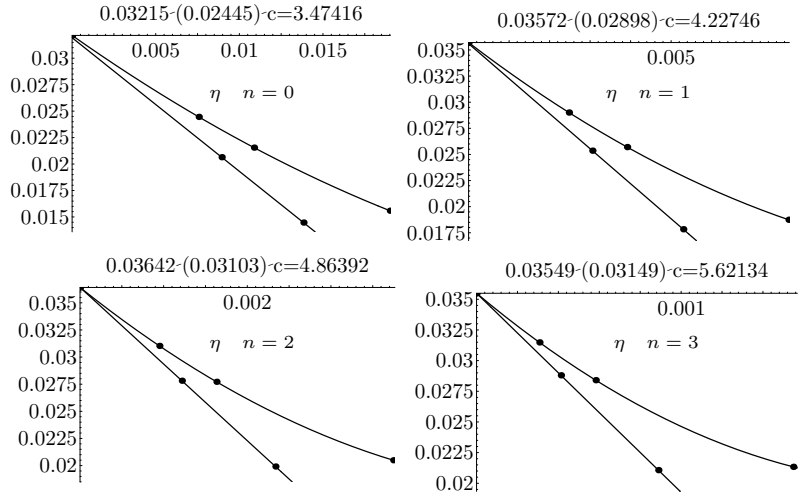

~\\[-4mm] \input cret0g.tps  ~\\~\\
\caption[]{Strong-coupling values
for the critical exponent $  \eta $
obtained from the expansion (\protect\ref{@eta})
via formula (\ref{coeffb}).
for increasing orders
$N=3,\dots,7$
 of the approximation.
The exponents are plotted against $x_N=e^{-cN^{1- \omega }}$.
Even approximants are connected by straight line and odd approximats
by slightly curved parabolas,
whose common intersection determines the critical exponents
expected for $N=\infty$.
}
\label{eta0g}\end{figure}

For the critical exponent
$ \eta $ we cannot use the same extrapolation procedure
since the expansion
(\ref{@eta}) starts out with $\bar g_0^2$,
so that there exists only an odd number of approximants
$ \eta _N$.
We therefore use two
alternative extrapolation procedures.
In the first we connect the even approximants $ \eta _2$ and $ \eta _4$
by a straight line and the odd ones
$ \eta _3, \eta _5, \eta _7$ by a slightly curved parabola,
and vary $c$ until there is an  intersection at $x=0$.
This yields the critical exponents $ \eta $
shown in Figs.~\ref{eta0g}.
Allowing for the inaccurate knowledge of $ \omega $, the
results may be stated as
\begin{equation}
 \eta_7
 =\left\{
\begin{array}{l}
0.03215+0.1327\times  ( \omega -0.810) \\
0.03572+0.0864 \times  ( \omega -0.805)\\
0.03642+0.0655 \times  ( \omega -0.800)\\
0.03549+0.0320\times  ( \omega -0.797)
\end{array}\right\}
~~~~~
{{\rm for}}       ~~~~~
\left\{ \begin{array}{l}
n=0\\
n=1\\
n=2\\
n=3
\end{array}\right\} ,%
\label{@etax}\end{equation}

\begin{figure}[tbhp]
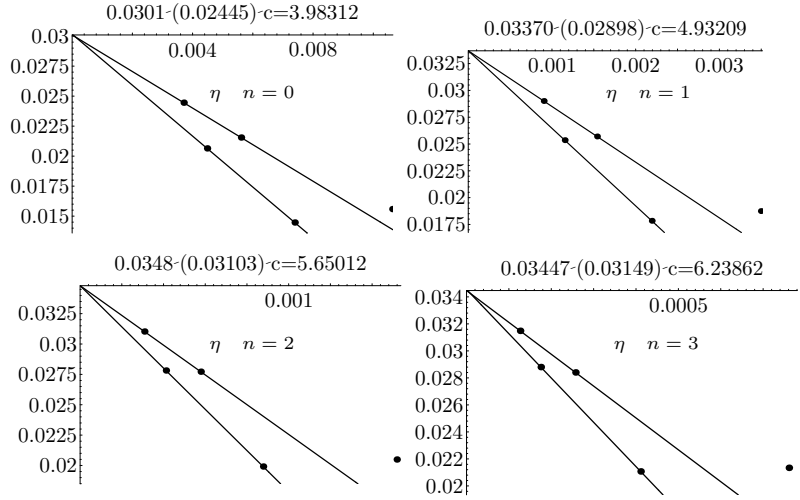

~\\ \input cret0h.tps  ~\\   ~\\
\caption[]{Plot analogous to Fig.~\protect\ref{eta0g}
but the extrapolation is found from the intersection of the straight lines
connecting
the last two even and odd approximants.
The resulting critical exponents
differ only little from those obtained in Fig.~\protect\ref{eta0g},
the differences given an estimate for the systematic error
of our results.
}
\label{eta0h}\end{figure}

Alternatively, we connect the last odd approximants
$ \eta _5$ and $\eta _7$ also by a straight line
and choose $c$ to make the lines intersect at $x=0$.
This yields the exponents
\begin{equation}
 \eta  _7
 =\left\{
\begin{array}{l}
0.03010+0.08760\times  ( \omega -0.810) \\
0.03370+0.03816 \times  ( \omega -0.805)\\
0.03480+0.01560 \times  ( \omega -0.800)\\
0.03447+0.00588\times  ( \omega -0.797)
\end{array}\right\}
~~~~~
{{\rm for}}       ~~~~~
\left\{ \begin{array}{l}
n=0\\
n=1\\
n=2\\
n=3
\end{array}\right\} ,%
\label{@etax22}\end{equation}
as
shown in Figs.~\ref{eta0h}, the $ \omega $ dependences
being somewhat weaker than in (\ref{@etax}).

Combining the two results
and using the difference to estimate the systematic error
of the extrapolation procedure,
we obtain for $ \eta $ the values
\begin{equation}
 \eta_7
 =\left\{
\begin{array}{l}
0.0311\pm0.001\\
0.0347\pm0.001\\
0.0356\pm0.001\\
0.0350\pm0.001
\end{array}\right\}
~~~~~
{{\rm for}}       ~~~~~
\left\{ \begin{array}{l}
n=0\\
n=1\\
n=2\\
n=3
\end{array}\right\} ,%
\label{@etax2}\end{equation}
whose $ \omega $-dependence is the average of
that in (\ref{@etax}) and
(\ref{@etax22}).

For our extrapolation procedure,
the power series for the critical exponent $ \gamma = \nu (2- \eta )$
are actually
better suited
than those for $ \eta $, since they
possess three even and three odd approximants, just as
$ \nu ^{-1}$. Advantages of this expansion have been
observed before \cite{remark}.
\begin{figure}[tbhp]
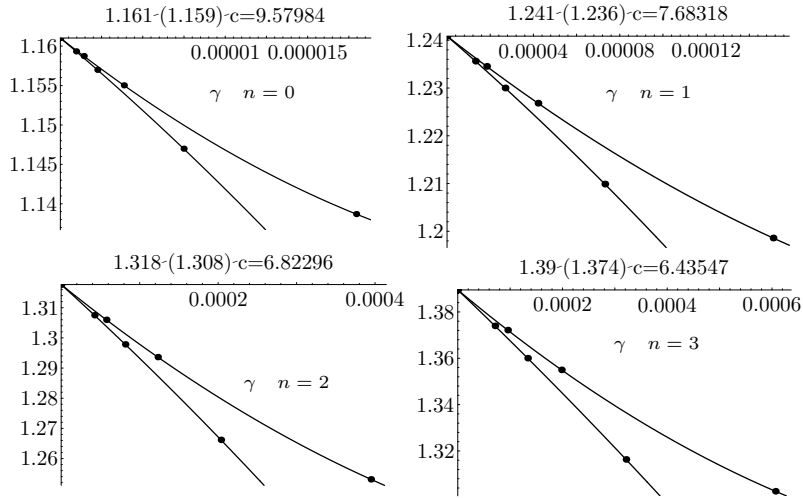

~\\ \input crga0g.tps   ~\\~\\
\caption[]{Strong-coupling values
for the critical exponent $ \gamma = \nu (2-\eta)=(2- \eta )/(2- \eta _m)$
obtained from a combination of the  expansions
 (\protect\ref{@eta})
and  (\protect\ref{@gammam})
via formula (\ref{coeffb}).
for increasing orders
$N=2,3,\dots,7$
 of the approximation.
The exponents are plotted against the variable $x_N=e^{-cN^{1- \omega }}$
and should lie on a straight line in the limit of large $N$.
Even and odd approximants are connected by slightly curved parabolas
whose common intersection determines the critical exponents
expected for $N=\infty$. The determination of the
constant $c$ is described in the text.
}
\label{ga0}\end{figure}
The associated plots are shown in Fig.~\ref{ga0}.
The extrapolated exponents are, including the $ \omega $-dependence,
\begin{equation}
 \gamma_7
 =\left\{
\begin{array}{r}
1.161-0.049\times ( \omega -0.810)\\
1.241-0.063\times ( \omega -0.805)\\
1.318-0.044\times ( \omega -0.800)\\
1.390-0.120\times ( \omega -0.797)
\end{array}\right\}
~~~~~
{{\rm for}}       ~~~~~
\left\{ \begin{array}{l}
n=0\\
n=1\\
n=2\\
n=3
\end{array}\right\} .%
\label{@gaex}\end{equation}
Unfortunately, the exponent $  \gamma = \nu (2- \eta ) $
is not very insensitive to
$ \eta $ since this is small compared to $2$, so that the extrapolation results
(\ref{@etax}) are more reliable than those obtained from $ \gamma $
via the scaling relation $\eta=2- \gamma / \nu $.
By combining
(\ref{@nuexdel})
and (\ref{@etax}), we find
from  $ \gamma = \nu (2- \eta )$:
\begin{equation}
 \gamma_7
 =\left\{
\begin{array}{r}
1.1589\\
1.2403\\
1.3187\\
1.3932
\end{array}\right\}
~~~~~
{{\rm for}}       ~~~~~
\left\{ \begin{array}{l}
n=0\\
n=1\\
n=2\\
n=3
\end{array}\right\} ,%
\label{@ga2ex}\end{equation}
the difference with respect to (\ref{@gaex})
showing the typical small errors of
our approximation, which are of the same
order as those
of the exponents obtained
in Ref.~\cite{GZ}.
As mentioned in the beginning,
the knowledge of the
large-order behavior
does not help to improve significantly the accuracy of the approximation.
In our theory, the most important exploited
information is the
knowledge of the
exponentially fast convergence which leads
to a linear behavior
of the resummation results of order $N$
in a  plot against
 $x_N=e^{-cN^{1- \omega }}$.
This knowledge,
which allows us to extrapolate our approximations for
$N=2,3,4,5,6,7$ quite well to
infinite order $N$,
seems to be more powerful than the knowledge of the large-order behavior
exploited by other authors (quoted in Table~\ref{TableI}).

The complete updated list of exponents
is shown in Table~\ref{TableI}, which also contains values
for the other critical exponents
$ \alpha =2- D \nu  $ and $ \beta = \nu (D-2+ \eta )/2$.

\begin{figure}[b]
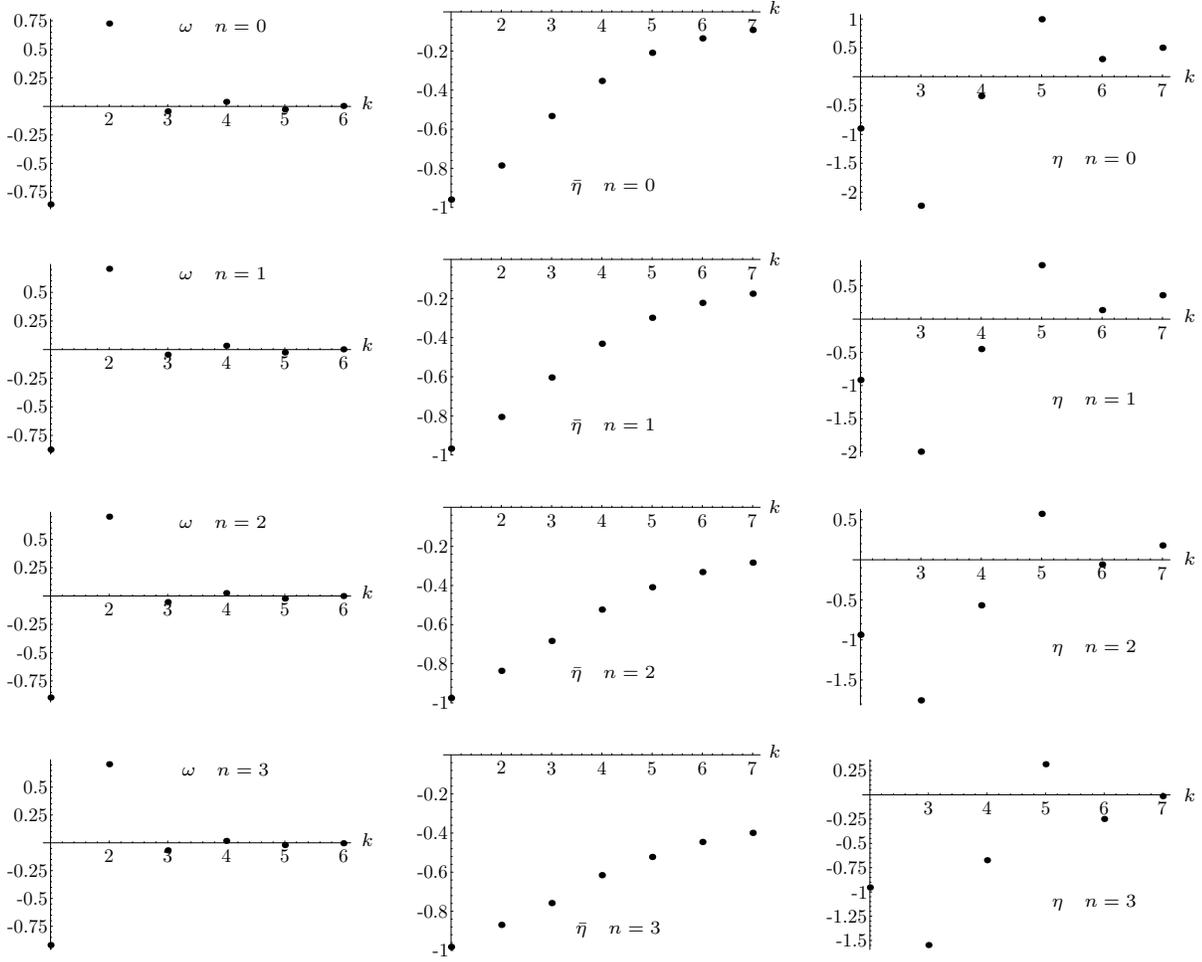

~\\
~~~~~~\input omerasr0.tps \\
~~~~~~\input omerasr1.tps  \\
~~~~~~\input omerasr2.tps    \\
~~~~~~\input omerasr3.tps      \\
\caption[Relative errors in predicting the
$k$th expansion coefficient
by fitting
the strong-coupling expansions
(\protect\ref{@omas})--(\protect\ref{@etpas})
for
$ \omega $, $\bar  \eta \equiv \nu^{-1}+ \eta -2 $]{Relative errors in predicting the
$k$th expansion coefficient
by fitting
the strong-coupling expansions
(\protect\ref{@omas})--(\protect\ref{@etpas})
for
$ \omega $, $\bar  \eta \equiv \nu^{-1}+ \eta -2 $, and $ \eta $
to the first $k-1$ expansion coefficients.
}
\label{errorsext}\end{figure}

~\\
\noindent
{\bf 5.} Let us now show that the large-order information
is indeed rather irrelevant to the critical exponents
within strong-coupling theory.
For this purpose we choose the coefficients
$c^{(i)}$ in the asymptotic formulas
 (\ref{@omas})--(\ref{@etpas})
to fit exactly the six known expansion coefficients
of $ \omega (\bar g)$ and the seven of $ \bar{\eta} (\bar g)$ and $ \eta (\bar g)$.
The coefficients are listed in Table \ref{ometetp},
and the associated fits are shown in Figs.~\ref{lo0}.
Since even and odd coefficients $ \eta ^{(k)}$
lie on twoseparate smooth  curves, we fit
the two sets separately.
These fits permit us to extend the presently available coefficients
and predict the results of future higher-loop calculations,
listed in Table \ref{@taextendec} up tp order 25.
The errors in these predictions are expected to be smallest for $ \omega ^{(k)}$,
as
illustrated in Figs.~\ref{errorsext}.

At this place we observe an interesting phenomenon:
According to Table~\ref{@taextendec}, the expansion coefficients
$ \omega ^{(k)}$  of $ \omega(\bar g) $ have alternating signs
and grow rapidly,
reaching precociously their asymptotic form
(\ref{@omas}), as we have seen in Figs.~\ref{lo0}.
Now, from $ \omega (\bar g)$ we can derive the so called $ \beta $-function
$ \beta (\bar g)\equiv \int d\bar g\,\omega (\bar g) $, and from this the expansion for the
bare coupling constant $\bar g_0(\bar g)=- \int d\bar g/ \beta (\bar g)$,
with coefficients
$\bar g_{0}^{(k)}$ listed in Table \ref{@taextendec2}.
From the standard instanton analysis \cite{PI},
we know that
the function
$\bar g_{0}(\bar g)$ has the same left-hand cut in the complex
$\bar g$-plane as the functions $ \omega (\bar g),\bar \eta(\bar g), \eta (\bar g)$,
with the same discontinuity proportional to
$e^{-{\rm const}/g}$ at the tip of the cut.
Hence the coefficients $\bar g_{0}^{(k)}$
must  have asymptotically a similar
alternating signs and a factorial growth.
Surprisingly, this expectation is not borne out by
the explicit seven-loop coefficients $\bar g_{0}^{(k)}$
following from (\ref{@omas}) in Table~\ref{@taextendec2}.
If we, however, look at the higher-order coefficients derived from  the
extrapolated $ \omega ^{(k)}$  sequence which are also listed
in that table,
we see that sign change and factorial growth
do eventually set in at the rather high order $11$. Before this order,
the coefficients
$\bar g_{0}^{(k)}$ look like those of a convergent series. Thus, if
 we would make a plot analogous  to those in Fig.~\ref{lo0}
for $\bar g_{0}^{(k)}$, we would observe
huge deviations from the asymptotic form up to an order much larger than 10.
In contrast, the inverse series
$\bar g(\bar g_{0})$  has expansion coefficients
 $\bar g_{k}$ which do approach rapidly their
asymptotic form, as seen in Table~\ref{@taextendec3}.
This is the reason why
our resummation of the critical exponents
$ \omega ,\bar  \eta , \eta$  as power series in  $\bar g_0$
yields good results already  at the available rather low order seven.

\begin{figure}[tbh]
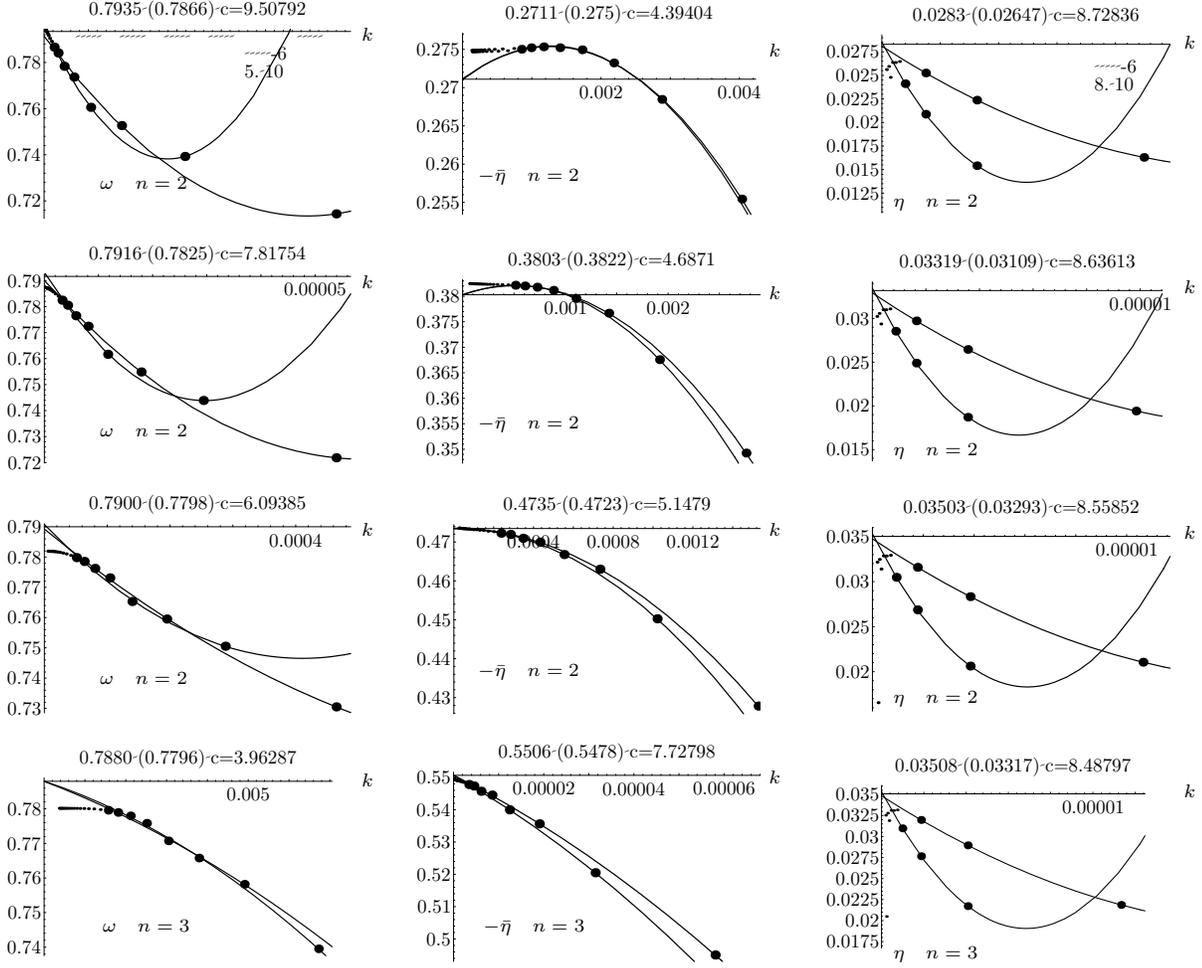

~\\
~~~~~~\input res250.tps \\
~~~~~~\input res251.tps \\
~~~~~~\input res252.tps \\
~~~~~~\input res253.tps \\
\caption[]{Extrapolation of resummed $ \omega,\bar \eta ,\eta $-values if
one $( \omega , \eta )$ or  two $(\bar  \eta )$ more expansion coefficients
of Table \ref{@taextendec}
are taken into account.
 The fat dots
show the resummed
values used for extrapolation, the small dots indicate
higher resummed values not used for the extrapolation.
The numbers on top specify the extrapolated values and the values of the last
approximation, corresponding to the leftmost fat dot.
}
\label{loext0}\end{figure}

Given the extrapolated list of expansion coefficients
in Table~\ref{@taextendec},
we may wonder how much these change the seven-loop results.
In  Figs.~\ref{loext0} we show the results. The known six-loops coefficients
of $ \omega(\bar g_0) $
and $ \eta(\bar g_0) $
 were extended by one extrapolated coefficient,
since this produces an even number of approximants
which can be most easily extrapolated to infinite order.
For $\bar \eta (\bar g_0)$ we use two more coefficients for the same
reason. The extrapolations are
shown in Figs.~(\ref{loext0}).
The resulting $ \omega _8$-values are lowered somewhat
with respect to $ \omega _6$
from
(\ref{@omex})
to
\begin{equation}
\omega_8
 =\left\{
\begin{array}{r}
0.7935\\ 0.7916\\ 0.7900\\ 0.7880
\end{array}\right\}
~~~~~
{{\rm for}}       ~~~~~
\left\{ \begin{array}{l}
n=0\\
n=1\\
n=2\\
n=3
\end{array}\right\}.~~ %
\label{@omexn}\end{equation}
The
new $ \eta $ values are
\begin{equation}
 \eta   _8
 =\left\{
\begin{array}{r}
0.02829-0.01675\times ( \omega -0.7935)\\
0.03319-0.01523\times ( \omega -0.7916)\\
0.03503-0.02428\times ( \omega -0.7900)\\
0.03537-0.01490\times ( \omega -0.7880)
\end{array}\right\}
~~~~~
{{\rm for}}       ~~~~~
\left\{ \begin{array}{l}
n=0\\
n=1\\
n=2\\
n=3
\end{array}\right\} ,
\label{@eta8}\end{equation}
lying reasonably close to the
previous seven-loop results
(\ref{@etax}), (\ref{@etax22}) for the smaller
$ \omega $-values (\ref{@omexn}).
The first set
yields
$ \eta_8
 =\left\{
0.0300, 0.0356, 0.0360,0.0354\right\} ,
$
the second
$ \eta_8
 =\left\{
0.0315 0.0342, 0.0349, .0345\right\} $.

For  $\bar  \eta $ we find the results
\begin{equation}
\bar  \eta _9
 =\left\{
\begin{array}{r}
-0.2711+0.0400\times ( \omega -0.810)\\
-0.3803+0.0974\times ( \omega -0.805)\\
-0.4735+0.1240\times ( \omega -0.800)\\
-0.5506+0.4761\times ( \omega -0.797)
\end{array}\right\}
~~~~~
{{\rm for}}       ~~~~~
\left\{ \begin{array}{l}
N=0\\
N=1\\
N=2\\
N=3
\end{array}\right\} .%
\label{@etapn}\end{equation}

\begin{figure}[t]
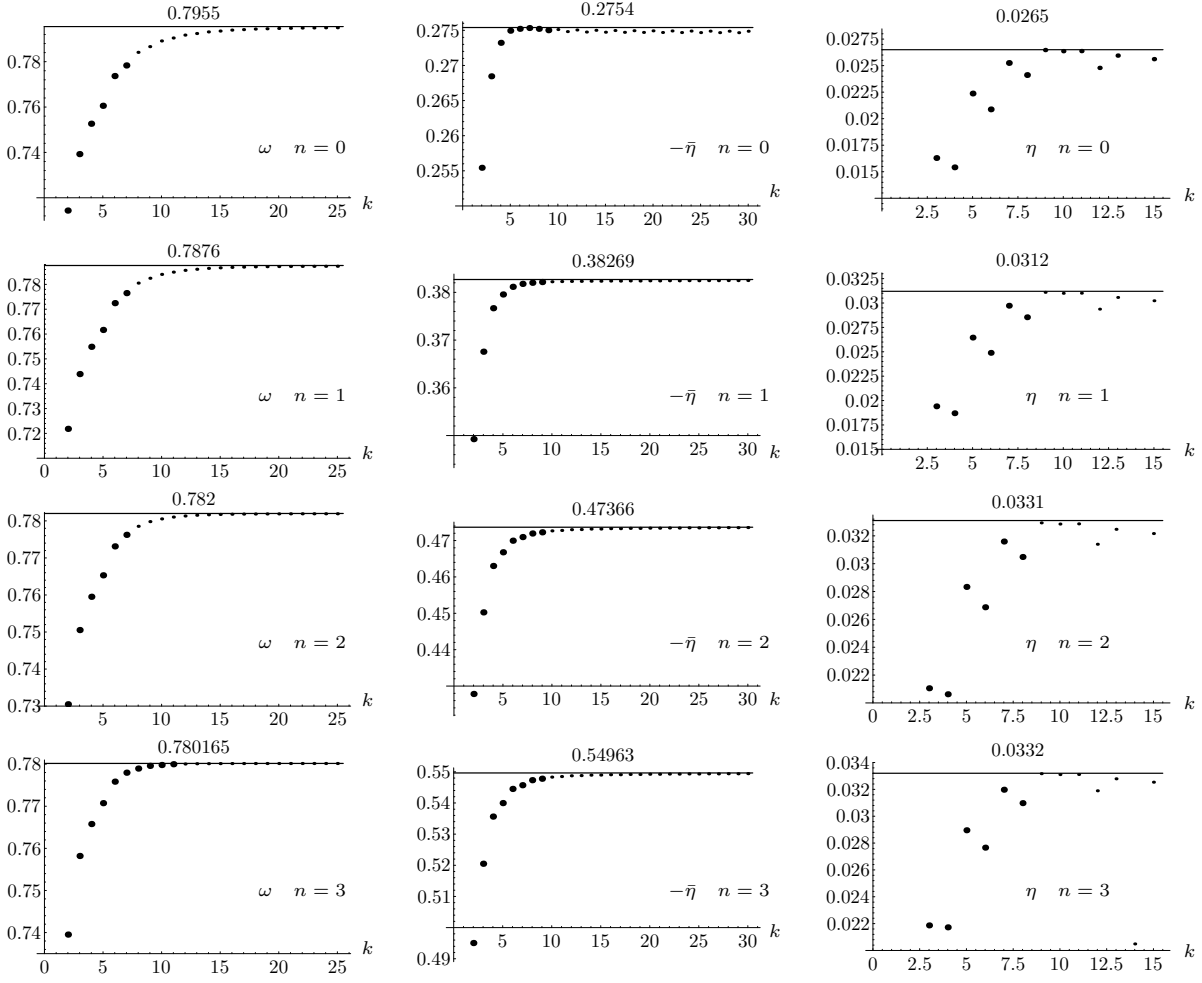

~\\
~~~~~~\input res25f0.tps \\
~~~~~~\input res25f1.tps \\
~~~~~~\input res25f2.tps \\
~~~~~~\input res25f3.tps \\
\caption[Direct plots
of the resummed $ \omega,\bar \eta ,\eta $-values
for all resummed values from all extrapolated expansion coefficients
of Table~\protect\ref{@taextendec}.
]{Direct plots
of the resummed $ \omega,\bar \eta ,\eta $-values
for all resummed values from all extrapolated expansion coefficients
of Table~\ref{@taextendec}.
The line is fitted to the maximum of all dots at the place
specified by the number on top. Fat and small dots distinguish the
resummed exponents used in the previous extrapolations
from the unused ones.
}
\label{loext0f}\end{figure}

It is interesting to observe how the
resummed values $ \omega _N,\bar  \eta _N, \eta _N$
obtained from the extrapolated expansion coefficients
in Table~\ref{@taextendec} continue to higher orders in $N$
This is shown in Figs.~\ref{loext0f}.
The dots converge against some specific values which, however, are
different from the
extrapolation results in Figs.~\ref{loext0} based on the
theoretical convergence behavior error $\approx e^{-cN^{1- \omega }}$.
We shall argue below that these results are  worse
than the properly extrapolated values.

\begin{figure}[tbhp]
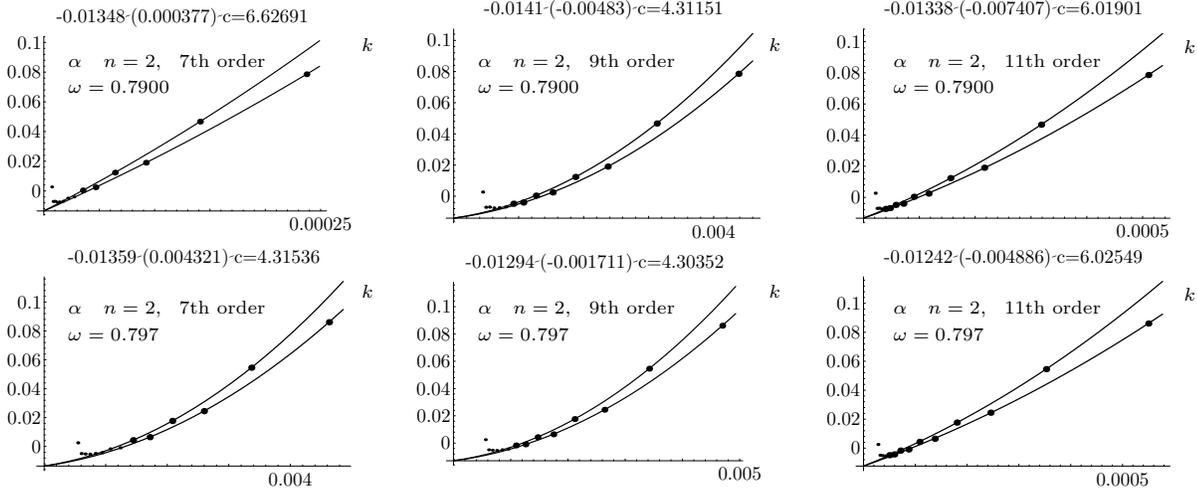

~\\
~~~~~~\input rsaln2.tps \\
~~~~~~\input rsal02.tps \\
\caption[]{Extrapolation of resummed $ \omega $-values if two more expansion coefficients
are taken from list in Table \ref{@taextendec}.
 The fat dots
show the resummed
values used for extrapolation, the small dots indicate
higher resummed values not used for the extrapolation.
}
\label{loalext}\end{figure}

All the above numbers agree
reasonably well with each other and with other estimates
in the literature listed in Table \ref{TableI}.
The only comparison with experiment
which is sensitive enough to judge the accuracy
of the results and the reliability of the resummation procedure
is provided by
the measurement of  $ \nu $ for $n=2$, where
the critical exponent
$ \alpha =2-3 \nu$ has been
extracted from the singularity $C\propto |1-T/T_c|^{- \alpha }$
in the specific heat at the $ \lambda $-point  of
superfluid helium
with high accuracy \cite{ahl}:
\begin{equation}
 \alpha =-0.01285\pm0.00038.
\label{@}\end{equation}
Since $  \nu  $ is of the order $2/3$, this measurement
is extremely sensitive to $ \nu $.
It is therefore useful do the resummations and extrapolations
for $N=2$
directly for the approximate $ \alpha $-values
 $ \alpha_N =2-3 \nu_N$,
once for  the six-loop
 $ \omega $-value $ \omega =0.8$, and once for a neighboring value
 $ \omega= 0.790$, to see the $ \omega $-dependence.
The results are shown in
Figs.~\ref{loalext}.
The extrapolated values for our $ \omega=0.8 $
in Table (\ref{TableI}) yield
\begin{equation}
 \alpha =-0.01294\pm0.00060,
\label{@}\end{equation}
in very good agreement with experiment.

The extrapolated expansion coefficients for  orders larger than 11
do not
carry significant information on the critical exponent $ \nu $.
The fact that the extrapolated expansion coefficients
should lie rather close to the true ones
as expected from the decreasing errors in
the plots in Fig.~(\ref{errorsext}) does not imply
the usefulness of the
new coefficients in Table~\ref{@taextendec} for obtaining better critical exponents.
The errors are only relatively small with respect to the
huge expansion coefficients.
The resummation procedure removes the  factorial growth
and becomes extremely sensitive to very small deviations
from thes huge coefficients.
This is the numerical consequence of the fact
discussed earlier that
the information residing in the exponentially small
imaginary part
of all critical exponents
near the
tip of the left-hand cut
in the complex $\bar g_0$-plane
has practically no effect upon the strong-coupling results at
infinite $\bar g_0$.

Note also that the critical exponents
which one would obtain from a
resummation of the extrapolated
expansion coefficients of high order
in Table~\ref{@taextendec} and their naive extrapolation
performed  in Figs.~\ref{loext0f} yield  a slightly worse result
for $ \alpha $ in superfluid helium.
Indeed, inserting $\bar  \eta =-0.47366$ and $ \eta =0.0331$
into the scaling relation $ \alpha =2-3/(2+\bar  \eta - \eta )$ we obtain
 $ \alpha =-0.0091$, which differs by about 25\% from
 the experimental $ \alpha $.

~\\~\\Acknowledgement\\
The author is grateful to F. Jasch and to
Drs. A. Pelster and V. Schulte-Frohlinde
for many useful discussions. He also thanks Drs. R. Guida
and A.I. Sokolov for sending him the
expansion coefficients of Ref.~\cite{MN}.
This research was supported in part by Deutsche Forschungsgemeinschaft (DFG).

 \newpage

\begin{table}[tbhp]
\caption[Fluctuation determinants and integrals over extremal field solution
]{Fluctuation
determinants and integrals over extremal field solution.\\ }
\label{Table 6}
%
\begin{tabular}{|l|llllll|}
$D$ & $D_L$ & $D_T$ & $I_1$ & $I_4$ & $I_6$ & $H_3$  \\
\hline
 3   &   10.544   $ \pm $ 0.004&   1.4571$\pm$ .0001  &     31.691522 &    75.589005  &	659.868352  &   13.563312 \\
 2   &   135.3  $\pm$ 0.1  &    1.465$\pm$ 0.001&    15.10965  &    23.40179
	 &     71.08023    &
	 9.99118  \\
\end{tabular}
\label{IHs}\end{table}
%
%
\begin{table}[tbhp]
\caption[Growth parameter of $D=3$ perturbation expansions of
   $\beta (\bar g),~\eta (\bar g)$ and $\eta ^{(4)}(\bar g)$]{Growth parameter
of $D=3$ perturbation expansions of
   $\beta (\bar g),~\eta (\bar g)$ and $\bar{\eta}= \eta + \nu ^{-1}-2$.\\ }
\label{Table: }
\footnotesize
 \begin{tabular}{|l|cccc|}
  %
    &  $n=0$ & $n=1$  & $n=2$ & $n=3$  \\
\hline
  $ a $      &    0.1662460   &    0.14777422 &    0.1329968 &  0.12090618\\
  $ b_  \omega   $ &   $4$   & ${9}/{2} $  & 9  &    ${11}/{2}$	       \\
  $  b _{\bar{\eta}}$ &   $3$   & ${7}/{2} $  & 4  &    ${9}/{2}$	       \\
   $b_\eta $ & 2 & ${5}/{2}$ & 3
& ${7}/{2}$ \\
   $10^2 \times  \gamma  $ &  8.5489(16)    &     3.9962(6)  &      1.6302(3)  &
	 0.59609(10)  \\
  $ 10^3 \times  \gamma  _{ \bar{\eta}}$ &
      10.107      &    6.2991   &     3.0836     &    1.2813  \\
   $10^3 \times  \gamma  _  \eta $ &
	2.8836     &    1.7972    &     0.8798   &      0.3656 \\
\end{tabular}
\label{compiled}
\end{table}
%

\begin{table}[tbhp]
\caption[Coefficients
of the large-order expansions (\protect\ref{@omas})--(\protect\ref{@etpas}), to fit
the known expansion coefficients of $ \omega $, $ \eta $, $ \bar{\eta}$.
The coefficients $ \eta^{(k)} $ possess two separate shorter  expansions
for even and odd $k$
]{Coefficients
of the large-order expansions (\protect\ref{@omas})--(\protect\ref{@etpas}), to fit
the known expansion coefficients of $ \omega $, $ \eta $, $ \bar{\eta}$.
The coefficients $ \eta^{(k)} $ possess two separate  expansions
for even and odd $k$.
 }
\label{Table asym}
%
\begin{tabular}{|l|l|rrrrrr|}
$$&$n$&$\hspace{-1cm}c^{(1)}~~~~~~~~~$ & $\hspace{-1cm}c ^{(2)}~~~~~~~~~$ & $\hspace{-1cm}c ^{(3)}~~~~~~~~~$ & $\hspace{-1cm}c ^{(4)}~~~~~~~~~$ & $\hspace{-1cm}c ^{(5)}~~~~~~~~~$& $\hspace{-1cm}c ^{(6)}~~~~~~~~~$  \\
\hline
$ \omega $&$0$  &0.2630147511231  &  3.440818282282    &-31.7673335904347 & 209.9430468590877&     -387.982076950413&   212.156573953838\\
$ $&$1$         &1.6353509905175  & -8.762940856111    & 32.5298724631003 &  49.5693979855620 &    -198.550118637547&  130.539359765928 \\
$ $&$2$         &4.1903240993241  &-32.521882201016    &159.2316083453243 &-271.5678237829086 &     185.521462986276 &  -36.220362093550\\
$ $&$3$         & 8.0659054235535 &-69.138003762384    &356.1987017927173 &-773.4084307341978 &     787.410568298674 & -297.863117806916\\
\hline
$  \bar{\eta}$&$0$  & ~15.4745287323349 & -263.105249597920  &~1695.85217994178   &-4797.25478881458  &~ 6198.21126891018& -2825.37877442787 \\
$ $&$1$  & ~10.9470420638543 & -169.697930580512   &~1074.82692242305   &-2886.57808941584  &~3577.48655305529 &-1577.19837665961\\
$ $&$2$  & ~~\,1.2481454871524 &  60.932456514040 &-409.59535356475    &~1526.62040773429  &-2300.49464074955&~ 1163.42553732492 \\
$ $&$3$  & -25.8032867124555 &  508.523659337565   &-3253.93912011988  &~9876.17157690861   &-13307.48621904672&~ 6257.59065449436 \\
\hline
$  \eta_{\rm o}$&$0$               & -6.3634296712273 &54.796985733992& -209.212694395258&159.7791383324933&& \\
$ $&$1$                            & -5.8608156341154 &58.173292227872& -237.158174423958&183.8456978302008&&  \\
$ $&$2$                            & -5.1086981057007 &64.465105150609& -285.116154230741&224.7597471858332&&  \\
$ $&$3$                            & -4.2039863427233 &76.269147128915& -364.452995945739&291.3878351595474&& \\
\hline
$  \eta_{\rm e}$&$0$  & -5.6929922203758& 15.551243915764&61.12469347544379&&&                            \\
$ $&$1$               & -5.3245881267711& 14.110708087849&81.2312043328075&&&                             \\
$ $&$2$               & -4.5203425601138&  9.799960635959&117.4131477198922&&&                             \\
$ $&$3$               & -3.1970976073075&  1.705210978430&176.4615812743069&&&                              \\
\end{tabular}
\label{ometetp}\end{table}

\begin{table}[tbhp]
\caption[]{Our seven-loop critical exponents (superscript s),
compared with results obtained by
other techniques. The
superscripts f and g refer to
other seven-loop expansions
in $D=3$ dimensions
(f $\in$ \cite{GZ},
g $\in$ \cite{MN}),
the other superscripts a--e
refer to six-loop results of a Pad\'e-Borel resummation
(a $\in$ \cite{anton},
(b $\in$ \cite{3},
c $\in$ \cite{4}),
and to
 five-loop expansions in $ \epsilon=4-D$
(d$\in$\cite{russ},
e$\in$\cite{13}).
For each of our results we give the highest approximation
before the extrapolated on
in parentheses. Only the first three rows and the $ \omega $-values
with superscript a in the entries for $n=0,1,2,3$ are new with respect
to the table in Ref.~\cite{kl}.}
\begin{tabular}{l|llllll|l}
$n$ & $g_c$ & $ \gamma ( \gamma_{6,7})$ & $ \eta (\eta_{6,7})$ & $ \nu( \nu_{6,7}) $ & $  \alpha $ & $  \beta $& $\omega$ ($\omega_{6,7}$)\\
\hline
0 &              & 1.161(1.159)$^{\rm s}$   & 0.0311$\pm0.001$$^{\rm s}$ & 0.5883(0.5864)$^{\rm s}$ & &      &0.810(0.773)\\[0mm]
& 1.413$\pm 0 .006^{\rm f}$&
1.160$\pm0.002 ^{\rm f} $ &
0.0284$\pm0.0025^{\rm f}$ &
0.5882$\pm0.0011^{\rm f}$&
0.235$\pm0.003^{\rm f}$ &
0.3025$\pm0.0008^{\rm f} $ &
0.812$\pm0.016^{\rm f}$\\[0mm]
 & 1.39$^{\rm g}$              & 1.1569$\pm0.0004 ^{\rm g} $ & 0.0297$\pm0.0009^{\rm g}$ & 0.5872$\pm0.0004^{\rm g}$&&&\\[0mm]
 & 1.402$^{\rm a}$   & 1.160$^{\rm a}$& 0.034$^{\rm a}$& 0.589$^{\rm a}$& 0.231$^{\rm a}$  & 0.305$^{\rm a}$ &\\
  & 1.421$\pm0.004^{\rm b}$   & 1.161$\pm0.003^{\rm b}$ & 0.026$\pm0.026^{\rm b}$ & 0.588$\pm0.001^{\rm b}$ & 0.236$\pm0.004^{\rm b}$ & 0.302$\pm0.004^{\rm b}$ & $0.794\pm0.06^{\rm b}$\\
  &$ 1.421\pm 0.008^{\rm c}$ &$ 1.1615\pm 0.002^{\rm c}$ & $0.027\pm 0.004^{\rm c}$ & 0.5880$\pm 0.0015^{\rm c}$ &      &  0.3020$\pm 0.0015^{\rm c}$& $0.80\pm 0.04^{\rm c}$\\
  & &$ 1.160\pm 0.004^{\rm e}$ & $0.031\pm 0.003^{\rm e}$ & 0.5885$\pm 0.0025^{\rm e}$ &      &  0.3025$\pm 0.0025^{\rm e}$& $0.82\pm 0.04^{\rm e}$\\
\hline
 1&  & 1.241(1.236)$^{\rm s}$  &0.0347$\pm0.001$$^{\rm s}$   & 0.6305(0.6270)$^{\rm s}$ & &  &0.805(0.772)\\[0mm]
 & 1.411$\pm0.004^{\rm f}$  & 1.240$\pm0.001 ^{\rm f} $ & 0.0335$\pm0.0025^{\rm f}$ & 0.6304$\pm0.0013^{\rm f}$&0.109$\pm0.004^{\rm f}$ &0.3258$\pm0.0014^{\rm f} $ &0.799$\pm0.011^{\rm f}$\\[0mm]
 & 1.40$^{\rm g}$  & 1.2378$\pm0.0006 ^{\rm g} $ & 0.0355$\pm0.0009^{\rm g}$ & 0.6301$\pm0.0005^{\rm g}$&&&\\[0mm]
& 1.419$^{\rm a}$& 1.239$^{\rm a}$& 0.038$^{\rm a}$ & 0.631$^{\rm a}$ & 0.107$^{\rm a}$& 0.327$^{\rm a}$  &0.781$^{\rm a}$\\
  & 1.416$\pm 0.0015^{\rm b}$ & 1.241$\pm 0.004^{\rm b}$ & 0.031$\pm 0.011^{\rm b}$ & 0.630$\pm 0.002^{\rm b}$& $0.110\pm0.008^{\rm b}$&$ 0.324\pm0.06^{\rm b}$& $0.788\pm0.003^{\rm b}$  \\
  & 1.416$\pm$0.004$^{\rm c}$  & 1.2410$\pm$0.0020$^{\rm c}$ &  0.031$\pm$0.004$^{\rm c}$ & 0.6300$\pm$0.0015$^{\rm c}$ &   & 0.3250$\pm$0.0015$^{\rm c}$ &0.79$\pm$0.03$^{\rm c}$\\
  & &  & $0.035\pm 0.002^{\rm d}$ & 0.628$\pm 0.001^{\rm d}$ &      & & $0.80\pm 0.02^{\rm d}$\\
  & &$ 1.1239\pm 0.004^{\rm e}$ & $0.037\pm 0.003^{\rm e}$ & 0.6305$\pm 0.0025^{\rm e}$ &      &  0.3265$\pm 0.0025^{\rm e}$& $0.81\pm 0.04^{\rm e}$\\
\hline
2 &  & 1.318(1.306)$^{\rm s}$ & $0.0356\pm0.001$$^{\rm s}$  & 0.6710(0.6652)$^{\rm s}$ &  & &0.800(0.772)\\[0mm]
 &  1.403$\pm0.003^{\rm f} $  & 1.317$\pm0.002 ^{\rm f} $ & 0.0354$\pm0.0025^{\rm f}$ & 0.6703$\pm0.0013^{\rm f}$&-0.011$\pm0.004^{\rm f}$ &0.3470$\pm0.0014^{\rm f} $ &0.789$\pm0.011^{\rm f}$\\[0mm]
 &  1.40$^{\rm g} $  & 1.3178$\pm0.001 ^{\rm g} $ & 0.0377$\pm0.0006^{\rm g}$ & 0.6715$\pm0.0007^{\rm g}$&& &\\[0mm]
 & 1.408$^{\rm a} $& 1.315$^{\rm a}$ & 0.039$^{\rm a} $& 0.670$^{\rm a}$ & -0.010$^{\rm a}$& 0.348$^{\rm a}$&0.780$^{\rm a}$\\
   & 1.406$\pm$0.005$^{\rm b}$ & 1.316$\pm$0.009$^{\rm b}$ & 0.032$\pm$0.015$^{\rm b}$ & 0.669$\pm$0.003$^{\rm b}$ & -0.007$\pm$0.009$^{\rm b}$ & 0.346$\pm$0.009$^{\rm b}$&$0.78\pm0.01^{\rm b}$\\
  & 1.406$\pm$0.004$^{\rm c}$ & 1.3160$\pm$0.0025$^{\rm c}$ & 0.033$\pm$0.004$^{\rm c}$ & 0.6690$\pm$0.0020$^{\rm c}$   &     & 0.3455$\pm$0.002$^{\rm c}$ &0.78$\pm$0.025$^{\rm c}$\\
  & &  & $0.037\pm 0.002^{\rm d}$ & 0.665$\pm 0.001^{\rm d}$ &      & &$0.79\pm 0.02^{\rm d}$\\
  & &$ 1.315\pm 0.007^{\rm e}$ & $0.040\pm 0.003^{\rm e}$ & 0.671$\pm 0.005^{\rm e}$ &      &  0.3485$\pm 0.0035^{\rm e}$& $0.80\pm 0.04^{\rm e}$\\
\hline
 3 &  &1.390(1.374)$^{\rm s}$  &$0.0350\pm0.0005$$^{\rm s}$  & 0.7075(0.7004)$^{\rm s}$ &  & &0.797(0.776)\\[0mm]
 & 1.391$\pm0.004^{\rm f}$      & 1.390$\pm0.005 ^{\rm f} $ &0.0355$\pm0.0025^{\rm f}$ & 0.7073$\pm0.0030^{\rm f}$&-0.122$\pm0.009^{\rm f}$ &0.3662$\pm0.0025^{\rm f} $ &0.782$\pm0.0013^{\rm f}$\\[0mm]
 & 1.39 $^{\rm g} $     & 1.3926$\pm0.001^{\rm g} $ &  0.0374$\pm0.0004^{\rm g} $ & 0.7096$\pm0.0008^{\rm g}$&&&\\[0mm]
 & 1.392$^{\rm a}$ & 1.386$^{\rm a}$ & 0.038$^{\rm a}$ & 0.706$^{\rm a}$ & -0.117$^{\rm a}$ & 0.366$^{\rm a}$&0.780$^{\rm a}$\\
  & 1.392$\pm$0.009$^{\rm b}$ & 1.390$\pm$0.01$^{\rm b}$ &  0.031$\pm0.022^{\rm b}$ &  0.705$\pm$0.005$^{\rm b}$ & -0.115$\pm0.015^{\rm b}$ & 0.362$^{\rm b}$ &$0.78\pm0.02^{\rm b}$\\
  & 1.391 $\pm$0.004$^{\rm c}$ & 1.386$\pm$0.004$^{\rm c}$ & 0.033$\pm$0.004$^{\rm c}$ & 0.705$\pm$0.003$^{\rm c}$ &    &    0.3645$\pm$0.0025$^{\rm c}$&$0.78\pm0.02^{\rm c} $\\
  & &  & $0.037\pm 0.002^{\rm d}$ & 0.79$\pm 0.02^{\rm d}$ &      & & $0.79\pm 0.02^{\rm d}$\\
  & &$ 1.390\pm 0.010^{\rm e}$ & $0.040\pm 0.003^{\rm e}$ & 0.710$\pm 0.007^{\rm e}$ &      &  0.368$\pm 0.004^{\rm e}$& $0.79\pm 0.04^{\rm e}$\\
\hline
\hline
  ~4 &           & 1.451(1.433) & 0.031(0.0289) & 0.737(0.732)&& & 0.795(0.780)\\
 & 1.375$^{\rm a}$ & 1.449$^{\rm a}$ & 0.036$^{\rm a}$& 0.738$^{\rm a}$ & -0.213$^{\rm a}$& 0.382$^{\rm a}$& 0.783$^{\rm a}$\\
\hline
 ~5 &  &1.511(1.487)  & 0.0295(0.0283)& 0.767(0.760) &  & &0.795(0.785)\\
 & 1.357$^{\rm a}$& 1.506$^{\rm a}$& 0.034$^{\rm a}$ & 0.766$^{\rm a}$ & -0.297$^{\rm a}$& 0.396$^{\rm a}$&0.788$^{\rm a}$\\
\hline
 ~6 &  &1.558(1.535) & 0.0276(0.0273) & 0.790(0.785) &  &  &0.797(0.792)\\
 & 1.339$^{\rm a}$ & 1.556$^{\rm a}$ & 0.031$^{\rm a}$ & 0.790$^{\rm a}$& -0.370$^{\rm a}$& 0.407$^{\rm a}$ &0.793$^{\rm a}$\\
\hline
~7 & & 1.599(1.577) & 0.0262(0.0260) & 0.810(0.807)&  &  &0.802(0.800)\\
 & 1.321$^{\rm a}$ & 1.599$^{\rm a}$& 0.029$^{\rm a}$& 0.811$^{\rm a}$ & -0.434$^{\rm a}$ & 0.417$^{\rm a}$&0.800$^{\rm a}$\\
\hline
~8 & & 1.638(1.612) & 0.0247(0.0246) & 0.829(0.825) &  & & 0.810(0.808)\\
 & 1.305$^{\rm a}$ & 1.637$^{\rm a}$ & 0.027$^{\rm a}$& 0.830$^{\rm a}$ & -0.489$^{\rm a}$ & 0.426$^{\rm a}$&0.808$^{\rm a}$\\
\hline
~9 &  & 1.680(1.643) & 0.0233(0.0233) & 0.850(0.841)&  & &0.817(0.815)
\\
 & 1.289$^{\rm a}$ & 1.669$^{\rm a}$& 0.025$^{\rm a}$ & 0.845$^{\rm a}$ & -0.536$^{\rm a}$ & 0.433$^{\rm a}$&0.815$^{\rm a}$\\
\hline
10 &  &1.713(1.670)& 0.0216(0.0220) & 0.866(0.854) &  & &0.824(0.822)\\
 & 1.275$^{\rm a}$& 1.697$^{\rm a}$& 0.024$^{\rm a}$ & 0.859$^{\rm a}$ & -0.576$^{\rm a}$ & 0.440$^{\rm a}$&0.822$^{\rm a}$\\
\hline
12 &  &1.763(1.716) & 0.0190(0.0198) & 0.890(0.877) &  & &0.838(0.835)\\
 & 1.249$^{\rm a}$& 1.743$^{\rm a}$& 0.021$^{\rm a}$ & 0.881$^{\rm a}$ & -0.643$^{\rm a}$ & 0.450$^{\rm a}$ &0.836$^{\rm a}$\\
\hline
14 & &1.795(1.750) & 0.0169(0.0178) & 0.905(0.894) &  & &0.851(0.849)\\
 & 1.227$^{\rm a}$& 1.779$^{\rm a}$& 0.019$^{\rm a}$ & 0.898$^{\rm a}$ & -0.693$^{\rm a}$& 0.457$^{\rm a}$&0.849$^{\rm a}$\\
\hline
16 &  & 1.822(1.779) & 0.0152(0.0161) & 0.918(0.907) &  & &0.862(0.860)\\
 & 1.208$^{\rm a}$ & 1.807$^{\rm a}$& 0.017$^{\rm a}$& 0.911$^{\rm a}$ & -0.732$^{\rm a}$& 0.463$^{\rm a}$&0.861$^{\rm a}$\\
18 & &1.845(1.803) & 0.0148(0.0137) & 0.929(0.918) && &0.873(0.869)\\
 & 1.191$^{\rm a}$ & 1.829$^{\rm a}$ & 0.015$^{\rm a}$ & 0.921$^{\rm a}$ & -0.764$^{\rm a}$ & 0.468$^{\rm a}$&0.871$^{\rm a}$\\
\hline
20 &        &1.864(1.822) & 0.0125(0.0135) & 0.938(0.927)& &&0.883(0.878)\\
 & 1.177$^{\rm a}$ & 1.847$^{\rm a}$ & 0.014$^{\rm a}$ & 0.930$^{\rm a}$ & -0.789$^{\rm a}$ & 0.471$^{\rm a}$&0.880$^{\rm a}$\\
\hline
24 & &1.890(1.850) & 0.0106(0.0116) & 0.950(0.939) & &&0.900(0.894)\\
 & 1.154$^{\rm a}$& 1.874$^{\rm a}$ & 0.012$^{\rm a}$ & 0.942$^{\rm a}$ & -0.827$^{\rm a}$ & 0.477$^{\rm a}$&0.896$^{\rm a}$\\
\hline
28 && 1.909(1.871)& 0.009232(0.01010)& 0.959(0.949) & &&0.913(0.906)\\
 & 1.136$^{\rm a}$ & 1.893$^{\rm a}$& 0.010$^{\rm a}$& 0.951$^{\rm a}$ & -0.854$^{\rm a}$ & 0.481$^{\rm a}$&0.909$^{\rm a}$\\
\hline
32 && 1.920(1.887) & 0.00814(0.00895)& 0.964(0.955)& &&0.924(0.915)\\
 & 1.122$^{\rm a}$ & 1.908$^{\rm a}$ & 0.009$^{\rm a}$& 0.958$^{\rm a}$& -0.875$^{\rm a}$& 0.483$^{\rm a}$&0.919$^{\rm a}$\\
\end{tabular}
\label{TableI}\end{table}

\begin{table}[p]
\caption[Coefficients of  extended perturbation expansions
obtained from  the large-order expansions
(\protect\ref{@omas})--(\protect\ref{@etpas})
for
 $ \omega $, $ \eta_m $, $ \eta $ up to $g^{12}$.]{Coefficients of  extended perturbation expansions
obtained from  the large-order expansions
(\protect\ref{@omas})--(\protect\ref{@etpas})
for
 $ \omega $, $ \eta_m $, $ \eta $ up to $g^{12}$.}
\scriptsize\begin{tabular}{|cc|ccccc|}
 $\!\!\!\!\!\!\!\!\!\! \!\!\!\!\!\!\!\!\!\!\!\!\!\!\!\!\!\!\!\!$&$k$&$n=0$&                  $n= 1$&                 $n=2$&                          $n=3$&                       \\
 \hline
$\!\!\!\!\!\!\!\!\!\! \omega ^{(k)}\!\!\!\!\!\!\!\!\!\!\!\!\!\!\!\!\!\!\!\!$&$0$   & {-1}~~&                       {-1}~~&                     {-1}~~&                  {-1}~~&                     \\
&$1$   & {2}&                       {2}&                     {2}&                  {2}&                     \\
&$2$   & {-95/72}&                  {-308/243}&              {-272/225}&       {-1252/1089}&            \\
&$3$   & {1.559690758}&       {1.404278391}&          {1.259667768}&     {1.131786725}&     \\
&$4$  & {-2.236580484}&      {-1.882634142}&         {-1.589642400}&    {-1.351666500}&    \\
&$5$  & { 3.803133000}&       {2.973285060}&          {2.346615000}&     {1.875335400}&             \\
&$6$  & {-7.244496000}&      {-5.247823000}&         {-3.867143000}&    {-2.904027000}&    \\
&$7$ & { 15.0706772}&        {10.0938530}&            {6.9384728}&       {4.8954471}&     \\
&$8$&  {-33.8354460}&       {-20.9045761}&          {-13.3833570}&      {-8.8630280}&     \\
&$9$&    {  81.4263429}&     {46.2983010}&           {27.5543342}&      {17.1018561}&     \\
&${10}$&{ -209.0371337}&  { -109.1428445}&         { -60.2679848}&    { -34.9985085}&   \\
&${11}$& { 570.2558985}&   { 272.8574773}&         { 139.5403648}&     { 75.6925030}&    \\
&${12}$&$   -1647.63898          $&$    -721.159283              $&$    -340.986931    $&$ -172.506443           $&   \\
&${13}$&$    5027.12671          $&$    2009.473994          $&$         877.142753     $&$ 413.269514           $&   \\
&${14}$&$  -16154.2792           $&$   -5888.53514         $&$         -2369.63316     $&$-1038.433113          $&   \\
&${15}$&$   54539.7867           $&$   18105.83253           $&$        6708.76515     $&$ 2731.28823           $&   \\
&${16}$&$ -193034.402            $&$  -58292.0930           $&$       -19865.5739      $&$-7505.78230          $&   \\
&${17}$&$  714771.195            $&$  196130.5369            $&$       61414.0151      $&$    21513.8526           $&   \\
&${18}$&$      ~~\,2.7637289\times10^6   $&$ -688418.829            $&$ -197883.530        $&$ -64215.5872           $&   \\
&${19}$&$      ~~\,1.1139530\times10^7   $&$      ~~\, 2.5166119\times10^6  $&$  663509.086$        &$ 199303.824            $&   \\
&${20}$&$      -4.6728706\times10^7   $&$         -9.5668866\times10^6   $&$   -2.3117713\times10^6$&$-642301.398           $&   \\
&${21}$&$      ~~\,2.0370346\times10^8   $&$      ~~\, 3.7765630\times10^7  $&$~~\,  8.3581769\times10^6 $&$ ~~\, 2.1465643\times10^6 $&   \\
&${22}$&$      -9.2152712\times10^8   $&$         -1.5460268\times10^8  $&$    -3.1317941\times10^7$&$       -7.4301887\times10^6 $&   \\
&${23}$&$      ~~\,4.3206669\times10^9   $&$      ~~\, 6.5552885\times10^8  $&$~~\,  1.2147059\times10^8$&$ ~~\,  2.6607749\times10^7 $&   \\
&${24}$&$      -2.0970132\times10^{10}$&$         -2.8755155\times10^9  $&$    -4.8714560\times10^8$&$       -9.8469119\times10^7  $&   \\
&${25}$&$      ~~\,1.0523676\times10^{11}$&$      ~~\, 1.3035111\times10^10 $&$~~\,  2.0178960\times10^9$&$ ~~\,  3.7621336\times10^8 $&   \\
\hline
$\!\!\!\!\!\!\!\!\!\!\bar  \eta^{(k)}\!\!\!\!\!\!\!\!\!\!\!\!\!\!\!\!\!\!\!\!$ &${1}$    &{-1/4}&             {-1/3}  &        {-2/5}&             {-5/11}&                       \\
&${2}$   &{1/16}&              {2/27}  &        {2/25}&            {10/121}&                   \\
&${3}$   &{-0.0357672729}&-0.0443102531&-0.0495134446 & -0.0525519564 &         \\
&${4}$   &{ 0.0343748465}& 0.0395195688& 0.0407881055&   0.0399640005 &        \\
&${5}$   &{-0.0408958349}&-0.0444003474&-0.0437619509&  -0.0413219917 &         \\
&${6}$   &{ 0.0597050472}& 0.0603634414& 0.0555575703 &  0.0490929344 &        \\
&${7}$   &{-0.09928487} & -0.09324948&  -0.08041336 &   -0.06708630 &         \\
&${8}$   &{ 0.18143353} &  0.15857090&   0.12955711 &    0.10413882 &         \\
&${9}$   &{-0.35946458} & -0.29269274&  -0.22839265 &   -0.17925852 &          \\
&${{10}}$&{ 0.76759881} &  0.58218392&   0.43525523 &    0.33488318 &        \\
&${{11}}$&{-1.75999735} & -1.24181846&  -0.88911482 &   -0.66904757 &          \\
&${{12}}$&{ 4.31887516} &  2.82935836&   1.93487570 &    1.41644564 &\\
&${13}$&$ -11.3068155     $&$-6.86145603     $& $-4.46485563         $&$-3.15991301     $&   \\
&${14}$&$  31.4831400      $&$17.65348358      $&$ 10.8846651         $&$7.40110473       $&   \\
&${15}$&$ -92.9568675      $&$-48.04185493     $&$-27.9476939        $&$-18.1528875     $&   \\
&${16}$&$ 290.205144      $&$  137.9015950      $&$75.3808299         $&$46.5326521      $&   \\
&${17}$&$-955.369710      $&$ -416.4425396     $&$-213.088140        $&$-124.454143     $&   \\
&${18}$&$3308.08653      $&$  1319.8954890      $&$630.008039         $&$346.784997      $&   \\
&${19}$&$ -12019.6749     $&$-4380.9238169     $&$-1944.51060        $&$-1005.36571     $&   \\
&${20}$&$  45726.095      $&$ 15196.764595       $&$6254.75115        $&$3028.67211      $&   \\
&${21}$&$-181763.39     $&$  -54989.750148     $&$ -20934.4636       $&$-9469.48945     $&   \\
&${22}$&$ 753530.79      $&$  207207.59430      $&$   72800.2529      $&$30694.0685      $&   \\
&${23}$&$ -3.2522981\times10^6$&$-811759.779                   $&$-262684.705        $&$-103030.713     $&   \\
&${24}$&$  ~~\, 1.4590604\times10^7 $&$ ~~\,3.3014377\times10^6 $&$982242.312          $&$357779.77      $&   \\
&${25}$&$ -6.7936016\times10^7$&$     -1.3919848\times10^7$&$-3.8016399\times10^6     $&$-1.2840285\times10^6$&   \\
\hline
$\!\!\!\!\!\!\!\!\!\! \eta ^{(k)}\!\!\!\!\!\!\!\!\!\!\!\!\!\!\!\!\!\!\!\!$&$1$   &{0}&                         {0}&                      {0}&                    0&                          \\
&$2$   &{1/108}&                     {8/729 }&                  {8/675}&                 40/3267&                     \\
&$3$   &{ 0.0007713749}&              {0.0009142223}&    {0.0009873600}&    0.0010200000&        \\
&$4$   &{ 0.0015898706}&              {0.0017962229}&    {0.0018368107}&    0.0017919257&        \\
&$5$   &{-0.0006606149}&             {-0.0006536980}&   {-0.0005863264}&   -0.0005040977&      \\
&$6$   &{0.0014103421}&               {0.0013878101}&    {0.0012513930}&    0.0010883237&                \\
&$7$   &{-0.001901867}&              {-0.0016976941}&   {-0.001395129}&    -0.001111499&       \\
&$8$   &{ 0.003178395}&               {0.0026439888}&    {0.002043629}&     0.001544149&        \\
&$9$   &{-0.006456700}&              {-0.0049783320}&   {-0.003585593}&    -0.002532983&       \\
&${10}$&{ 0.012015200}&              { 0.0084255120}&   { 0.005570210}&     0.003647578&       \\
&${11}$&{-0.029656348}&              {-0.0194143738}&   {-0.012066168}&    -0.007451622&      \\
&${12}$&{ 0.064239639}&              { 0.0378738590}&   { 0.021403479}&     0.012148673&        \\
&${13}$&$-0.180415293$&$ -0.0992734993$&$-0.0527914785 $&       $ -0.0282931664$&   \\
&${14}$&$ 0.451904799$&$ ~0.2230420013$&$0.10748443321$&$ 0.0528085190 $&   \\
&${15}$&$-1.409286997$&$-0.6472476781$&$-0.2928360472$& $ -0.135567321 $&   \\
&${16}$&$ 4.021490037$&$1.65386975$&$0.67741438871$& $ 0.287414739  $&   \\
&${17}$&$-13.75881440$&$-5.24609037 $&    $-2.01071514 $&       $ -0.801301742 $&   \\
&${18}$&$ 44.09028452$&$15.0426293  $&    $5.21919799  $&       $ 1.907241838  $&   \\
&${19}$&$-164.205876 $&  $-51.7544723 $&    $-16.7458885 $&       $ -5.728643910 $&   \\
&${20}$&$ 583.728411  $&  $164.571258  $&    $48.2146655  $&       $ 15.13540671  $&   \\
&${21}$&$-2352.30706 $&  $-610.647520 $&    $-166.308263 $&       $ -48.72074256 $&   \\
&${22}$&$ 9182.66367  $&  $2131.95908  $&    $525.890013  $&       $ 141.4691142  $&   \\
&${23}$&$-39836.8326 $&  $-8491.50902 $&    $-1941.60261 $&       $ -486.0716246 $&   \\
&${24}$&$ 169338.243  $&  $32279.4193  $&    $6686.67654  $&       $ 1538.009823  $&   \\
&${25}$&$-787352.117 $&  $-137442.343 $&    $-26325.2747 $&       $ -5621.263980 $&   \\
\end{tabular}
\label{@taextendec}\end{table}

\begin{table}[p]
\caption[Coefficients of $\bar g_0(\bar g)$
obtained from
 extended perturbation expansions
obtained from  the large-order expansions
(\protect\ref{@omas})--(\protect\ref{@etpas})
for
 $ \omega (\bar g) $
 up to $g^{25}$]{Coefficients of $\bar g_0(\bar g)$
obtained from
 extended perturbation expansions
obtained from  the large-order expansions
(\protect\ref{@omas})--(\protect\ref{@etpas})
for
 $ \omega (\bar g) $
 up to $g^{25}$.}
\scriptsize\begin{tabular}{|cc|ccccc|}
 $\!\!\!\!\!\!\!\!\!\! \!\!\!\!\!\!\!\!\!\!\!\!\!\!\!\!\!\!\!\!$&$k$&$n=0$&                  $n= 1$&                 $n=2$&                          $n=3$&                       \\
 \hline
$\!\!\!\!\!\!\!\!\!\!\bar g_{0}^{(k)}\!\!\!\!\!\!\!\!\!\!\!\!\!\!\!\!\!\!\!\!$&$1$   & {1}~~&                       {1}~~&                     {1}~~&                  {1}~~&                     \\
&$2$   &$+ 1                     $& $+ 1                     $&  $+ 1                    $&$+ 1                     $\\
&$3$   &$+ 337/432               $& $+ 575/729               $&  $+ 539/675              $&$+ 2641/3267             $\\
&$4$   &$+ 0.61685694588 $& $+ 0.62411053351 $&  $+ 0.63484885720$&$+ 0.64721832545 $\\
&$5$   &$+ 0.44266705709 $& $+ 0.45557995443 $&  $+ 0.47149516705$&$+ 0.48876206059 $\\
&$6$   &$+ 0.35597494073 $& $+ 0.35927512536 $&  $+ 0.36876801981$&$+ 0.38195333853 $\\
&$7$   &$+ 0.21840619207 $& $+ 0.23668638696 $&  $+ 0.25507866294$&$+ 0.27372501773 $\\
&$8$   &$+ 0.23516444398 $& $+ 0.22010271935 $&  $+ 0.21833333377$&$+ 0.22423492600 $\\
&$9$   &$+ 0.02522653990 $& $+ 0.07797541233 $&  $+ 0.11146939079$&$+ 0.13619054953 $\\
&${10}$&$+ 0.32466738893 $& $+ 0.21722566733 $&  $+ 0.17071122132$&$+ 0.15281461436 $\\
&${11}$&$- 0.46084539160 $& $- 0.17781419227 $&  $- 0.04796874299$&$+ 0.01851106465 $\\
&${12}$&$+ 1.36111296151 $& $+ 0.62177013621 $&  $+ 0.32371445346$&$+ 0.19688967179 $\\
&${13}$&$- 3.42004319798 $& $- 1.33935153089 $&  $- 0.55625249070$&$- 0.23297770291 $\\
&${14}$&$+ 9.68597708110 $& $+ 3.55457753745 $&  $+ 1.44715002648$&$+ 0.65263956302 $\\
&${15}$&$- 28.5286709455 $& $- 9.51594412468 $&  $- 3.51833733708$&$- 1.41477489238 $\\
&${16}$&$+ 88.9376821020 $& $+ 27.1477264424 $&  $+ 9.31404148366$&$+ 3.53850316476 $\\
&${17}$&$- 291.235785543 $& $- 81.0609653416 $&  $- 25.6008150903$&$- 9.00262320492 $\\
&${18}$&$+ 1000.66241399 $& $+ 253.799830529 $&  $+ 73.8458792207$&$+ 24.1544067361 $\\
&${19}$&$- 3599.15484483 $& $- 830.784519325 $&  $- 222.359395181$&$- 67.4743858406 $\\
&${20}$&$+ 13526.5566605 $& $+ 2838.71379781 $&  $+ 698.348588943$&$+ 196.518945901 $\\
&${21}$&$- 53025.6841577 $& $- 10107.5962344 $&  $- 2283.46544025$&$- 595.358754523 $\\
&${22}$&$+ 216470.154554 $& $+ 37445.8720302 $&  $+ 7762.54138666$&$+ 1873.85881729 $\\
&${23}$&$- 918905.735057 $& $- 144134.115732 $&  $- 27396.7807350$&$- 6119.04352841 $\\
&${24}$&$+ 4050397.96349$&  $+ 575646.134976 $&  $+ 100259.282083$&$+ 20705.5670994 $\\
&${25}$&$- 18514433.0840$&  $- 2382463.70507 $&  $- 379975.758849$&$- 72517.4413857 $\\
\end{tabular}
\label{@taextendec2}\end{table}
\begin{table}[p]
\caption[Coefficients of $\bar g(\bar g_0)$
obtained from
 extended perturbation expansions
obtained from  the large-order expansions
(\protect\ref{@omas})--(\protect\ref{@etpas})
for
 $ \omega (\bar g) $
 up to $g^{25}$]{Coefficients of $\bar g(\bar g_0)$
obtained from
 extended perturbation expansions
obtained from  the large-order expansions
(\protect\ref{@omas})--(\protect\ref{@etpas})
for
 $ \omega (\bar g) $
 up to $g^{25}$.}
\scriptsize\begin{tabular}{|cc|ccccc|}
 $\!\!\!\!\!\!\!\!\!\! \!\!\!\!\!\!\!\!\!\!\!\!\!\!\!\!\!\!\!\!$&$k$&$n=0$&                  $n= 1$&                 $n=2$&                          $n=3$&                       \\
 \hline
$\!\!\!\!\!\!\!\!\!\!\bar g^{(k)}\!\!\!\!\!\!\!\!\!\!\!\!\!\!\!\!\!\!\!\!$&$1$   & {1}~~&                       {1}~~&                     {1}~~&                  {1}~~&                     \\
&$2$   &$- 1                           $& $- 1                     $&  $- 1                      $&$- 1                     $\\
&$3$   &$+ 527/432                     $& $+ 883/729               $&  $+ 811/675                $&$+ 3893/3267             $\\
&$4$   &$- 1.7163939829$& $- 1.680351960126292     $&  $- 1.642256264617284      $&$- 1.60528382897736      $\\
&$5$   &$+ 2.7021635328$& $+ 2.591685040643859     $&  $+ 2.481604560563785      $&$+ 2.378891143794822     $\\
&$6$   &$- 4.6723281932$& $- 4.363908063002809     $&  $- 4.073635397816119      $&$- 3.813515390028028     $\\
&$7$   &$+ 8.7648283753$& $+ 7.926093595753771     $&  $+ 7.180326093595318      $&$+ 6.539645290718699     $\\
&$8$   &$- 17.684135663$& $- 15.39841276963578     $&  $- 13.47981441366666      $&$- 11.90293506879397     $\\
&$9$   &$+ 38.129348202    $& $+ 31.80063328573243     $&  $+ 26.79259688548747      $&$+ 22.86325133485651     $\\
&${10}$&$- 87.419391225   $& $- 69.48420478282783     $&  $- 56.1279351033013       $&$- 46.14596304145893     $\\
&${11}$&$+ 212.28789113   $& $+ 160.0400066477353     $&  $+ 123.4985362910675      $&$+ 97.5437851896555      $\\
&${12}$&$- 544.33806227   $& $- 387.4479410496121     $&  $- 284.6297746951519      $&$- 215.3826650602743     $\\
&${13}$&$+ 1470.2445538   $& $+ 983.719405302971      $&  $+ 685.668309006505       $&$+ 495.7770927688912     $\\
&${14}$&$- 4175.1804881   $& $- 2614.933427024693     $&  $- 1723.672999416843      $&$- 1187.794187410145     $\\
&${15}$&$+ 12447.739474   $& $+ 7268.064649337187     $&  $+ 4516.120357408118      $&$+ 2958.336103932099     $\\
&${16}$&$- 38915.141370   $& $- 21101.49568383381     $&  $- 12320.85534817637      $&$- 7652.516371929849     $\\
&${17}$&$+ 127440.33105   $& $+ 63943.24392789235     $&  $+ 34975.98186824855      $&$+ 20545.02631707489     $\\
&${18}$&$- 436738.21140   $& $- 202094.1329180427     $&  $- 103252.1798678474      $&$- 57215.98372843337     $\\
&${19}$&$+ 1564637.2472  $& $+ 665710.523944826  $&  $+ 316810.7604431689      $&$+ 165210.8008728902     $\\
&${20}$&$- 5853354.4104  $& $- 2283830.09806744$& $   - 1009811.938755735 $&     $- 494409.476944406      $\\
&${21}$&$+ 22839087.694  $& $+ 8153184.95412866$&  $  + 3341698.327836095 $&     $+ 1532757.736028176$\\
&${22}$&$- 92830002.172   $& $- 30260412.9590709$&  $- 11473421.52345331 $&     $- 4920271.757368278$\\
&${23}$&$+ 392524311.64  $& $+ 116646023.338810   $&   $+ 40840739.00033049 $&     $+ 16345368.25243382$\\
&${24}$&$- 1724406456.3  $& $- 466498175.446816 $&   $- 150595276.6851763 $&     $- 56159032.51385756$\\
&${25}$&$+ 7860313710.5  $& $+ 1933471826.94197$&    $+ 574727529.9905997 $&     $+ 199417525.2243582$\\
\end{tabular}
\label{@taextendec3}\end{table}


\begin{thebibliography}{11}
\bibitem{not}
Note that the expansion for $ \eta _m=2- \nu ^{-1}$ in
Eq.~(61) of Ref.~\cite{kl} contained a printing error.The sign of
the $\hat g_0^2$-term
must be reversed.
\end{thebibliography}
\end{document}